\documentclass[aps,pra,reprint,superscriptaddress]{revtex4-1}
\usepackage{epsfig,amssymb,amsmath,amsthm,graphicx,color}
\def\ket#1{\lvert #1\rangle}

\begin{document}

\title{Photon molecules in atomic gases trapped near photonic crystal waveguides}

\author{James S.~Douglas}
\email{Corresponding author james.douglas@icfo.eu}
\affiliation{ICFO-Institut de Ciencies Fotoniques, The Barcelona Institute of Science and Technology, 08860 Castelldefels (Barcelona), Spain}
\author{Tommaso Caneva}
\affiliation{ICFO-Institut de Ciencies Fotoniques, The Barcelona Institute of Science and Technology, 08860 Castelldefels (Barcelona), Spain}
\author{Darrick E.~Chang}
\affiliation{ICFO-Institut de Ciencies Fotoniques, The Barcelona Institute of Science and Technology, 08860 Castelldefels (Barcelona), Spain}

\date{\today}

\begin{abstract}
Realizing systems that support robust, controlled interactions between individual photons is an exciting frontier of nonlinear optics. To this end, one approach that has emerged recently is to leverage atomic interactions to create strong and spatially non-local interactions between photons. In particular, effective interactions have been successfully created via interactions between atoms excited to Rydberg levels. Here, we investigate an alternative approach, in which atomic interactions arise via their common coupling to photonic crystal waveguides. This technique takes advantage of the ability to separately tailor the strength and range of interactions via the dispersion engineering of the structure itself, which can lead to qualitatively new types of phenomena. As an example, we discuss the formation of correlated transparency windows, in which photonic states of a certain number and shape selectively propagate through the system. Through this technique, we show in particular that one can create molecular-like potentials that lead to molecular bound states of photon pairs.
\end{abstract}

\maketitle

\section{Introduction}

In recent years, there has been tremendous progress to realize systems that are capable of achieving strong interactions between individual photons \cite{Chang2014a}. One common approach has been to couple single atoms (or other quantum emitters) to high-finesse optical cavities, to take advantage of the intrinsic nonlinear nature of these two-level systems \cite{Turchette1995a,Imamoglu1997c,Duan2004a,Birnbaum2005a,Haroche2006,Kimble2008a,Reiserer2015a}. More recently, gases of cold Rydberg atoms have been investigated \cite{Pritchard2010a,Dudin2012a,Pritchard2013a,Peyronel2012a,Baur2014a,Gorniaczyk2014a,Tiarks2014a}. In this case, the optical nonlinearities are effectively generated via strong atom-atom interactions, with a novel consequence being that the nonlinearity becomes spatially non-local in character \cite{Gorshkov2011a}. It has been experimentally shown that this type of nonlinearity can give rise to exotic states such as a two-photon bound state \cite{Firstenberg2013a}, and a number of other few- and many-body states of light have been theoretically predicted \cite{Otterbach2013a,Bienias2014a,Maghrebi2015a}.

\begin{figure}
\centering
\includegraphics[scale=0.94]{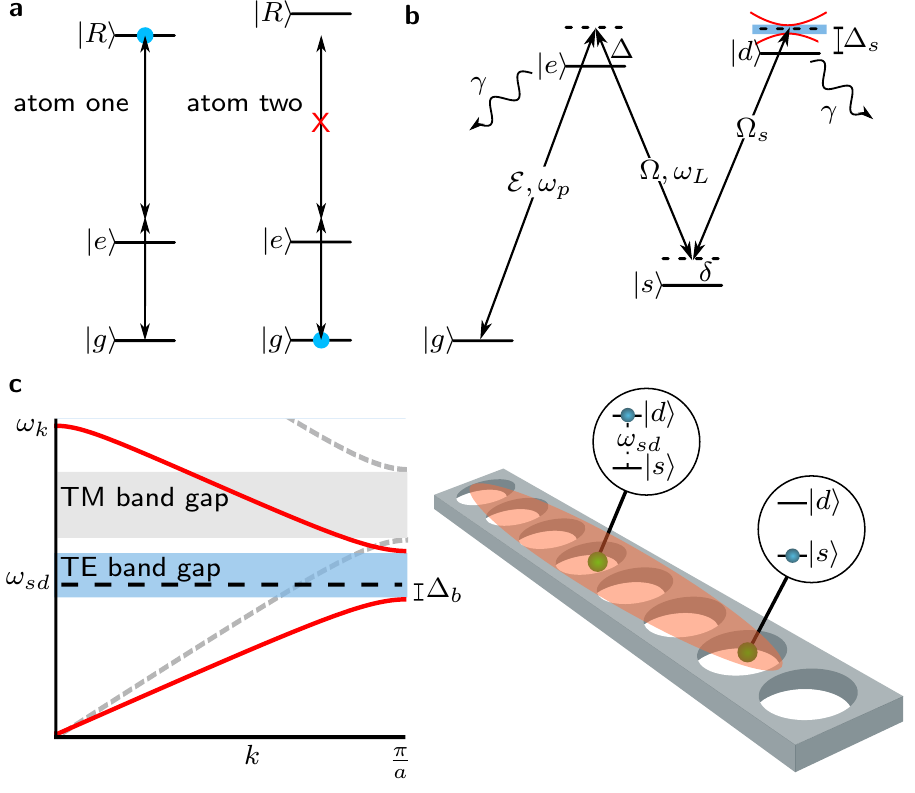}
\caption{(a) Rydberg interactions between atoms mediate photon-photon interactions. Here atom 1 is excited to Rydberg level $R$ by two-photon excitation and the Rydberg interaction shifts the $R$ level in other surrounding atoms by an amount much larger than the two-photon transition width, making the transition effectively two-level for a second photon entering the system. (b) Atoms with three internal levels support EIT. The input quantum probe field $\mathcal{E}$ couples to the transition $\ket{g}-\ket{e}$, while a second transition $\ket{s} - \ket{e}$ couples to a classical drive field.  By adding an additional level $\ket{d}$ that couples to the band edge modes of a photonic crystal and driving the transition  $\ket{s}-\ket{d}$ off-resonantly an effective interaction between the $\ket{s}$ levels can be generated. For simplicity we assume states $\ket{e}$ and $\ket{d}$ both have free-space spontaneous decay rate $\gamma$. (c) The periodic dielectric structure of a photonic crystal (right) leads to band structure (left) in the dispersion relation for photons propagating in the dielectric. A band gap occurs over frequencies in which propagation is forbidden due to strong interference from the underlying structure. For photonic crystals that support multiple modes, e.g., TE and TM (represented here by the solid and dotted lines), the band gaps of different modes may occur at different frequencies. The $\ket{s}-\ket{d}$ transition of the atoms (green spheres) trapped near the photonic crystal couples to the photonic crystal bands. When the transition frequency $\omega_{sd}$ is in the band gap (with detuning $\Delta_b$ from the band edge) atoms can couple with one-another via evanescent fields in the photonic crystal (illustrated in red), yielding a dipole-dipole interaction between the atoms. }
\label{fig:level_int}
\end{figure}

Key to this approach is that atoms excited to a Rydberg state, in particular through the absorption of a photon, exert a strong dispersive effect on proximal atoms, shifting their transition frequencies to Rydberg states by an amount proportional to $\sim 1/r^6$, with $r$ being the interatomic separation (see Fig.~\ref{fig:level_int}a) \cite{Saffman2010a}. This in turn, leads to a shift in the optical susceptibility of the atoms, which can be interpreted as a change of refractive index that depends on the number and position of photons in the system. In practice, the $r$ dependence of the Rydberg interaction presents somewhat of a constraint. Specifically, one must reach extremely high Rydberg states in order to induce a significant nonlocal effect ($n\sim 100$ in Ref.~\cite{Firstenberg2013a} to achieve a blockade radius of $18\mu$m). This in turn yields a large level shift ($\sim 50$GHz for atoms separated by $4\mu$m in Ref.~\cite{Firstenberg2013a})); however, all of the interesting variation in the atomic refractive index occurs over much smaller bandwidths of a few MHz (characteristic of the atomic linewidth). In this article, we demonstrate the new opportunities arising for nonlinear optics if the strength and range of the interaction can be independently adjusted. In particular, we show that the highly tunable nature of systems coupling atoms with nano-photonic devices can greatly extend the gamut of photon-photon interactions. 

Our work is inspired by recent developments to interface cold atoms with photonic crystals \cite{Thompson2013a,Goban2013a,Yu2013a,Tiecke2014a,Goban2015a}. It has been proposed that one can control the type and range of atomic interactions \cite{Kurizki1990a,John1991a,Shahmoon2013a,Douglas2014a,Tudela2014a}, and hence the interactions between photons \cite{Shahmoon2015a}, by engineering the underlying optical dispersion of these structures. We specifically study the dynamics of photons propagating through an atomic ensemble coupled to the waveguide under conditions of electromagnetically induced transparency (EIT), which occurs for three level atoms such as shown in Fig.~\ref{fig:level_int}a-b \cite{Fleischhauer2005a}. Without long-range interactions, EIT is an interference effect that allows an optical probe field to propagate without absorption due to the interference created by a second pump field when a ``two-photon resonance'' condition is satisfied, $\delta = 0$ in  Fig.~\ref{fig:level_int}b. As a linear optical effect, a probe pulse of any photon number and shape within a certain bandwidth propagates in this transparent manner.  

Atomic interactions (that for example shift level $\ket{s}$ in  Fig.~\ref{fig:level_int}b) shift the two-photon resonance, altering the propagation of photons in the system and effective photon-photon interactions result. Provided these interactions are sufficiently long-range and smooth, we show that one is able to create ``correlated transparent states,'' in which only pulses of a desired photon number and shape will propagate through. As an example, we provide an explicit construction to generate a two-photon molecule, in which two photons are bound at a fixed separation by an effective spring, and where ``phonon'' oscillations constitute the fundamental excitations of such a state. 

The article is structured as follows. In Section \ref{sec:phc_int}, we briefly review the physics governing the interaction between atoms and photonic crystal waveguides. In particular, we show how tunable long-range interactions between atoms emerge when the atomic transition frequency is situated within an optical ``band gap'' of the structure. In Section \ref{sec:eit_interations}, we show how EIT can be used to convert these atom-atom interactions into effective interactions between photons. In this regime we find an effective equation for photons propagating in the system that supports the presence of correlated transparent states. In Section \ref{sec:photon_molecules}, we show how the photon interactions can be tuned to create molecular-like states of photons and, in Section \ref{sec:tech_req}, 
we derive the conditions to observe such a state, demonstrating that the required parameters are within reach of state-of-the-art nano-photonic systems, before concluding in Section \ref{sec:summary}.


\section{Interactions mediated by photonic crystals}\label{sec:phc_int} 

Photonic crystals are dielectric structures in which the refractive index is modulated periodically \cite{Joannopoulos2008}. At some frequencies, light input into the dielectric reflects constructively from these modulations and the system acts like a mirror, preventing propagation. This leads to the presence of band gaps in the dispersion relation for the photonic modes, as shown in Fig.~\ref{fig:level_int}(c). When an atom trapped nearby the photonic crystal is excited at a frequency in the band gap, it is forbidden from emitting a propagating photon into the dielectric. However, it can generate an exponentially decaying evanescent field that forms a localized photonic cloud around the atom \cite{John1990a}. 
It has been formally shown that this photonic cloud has the same functionality as a real cavity of the same size \cite{Douglas2014a}, with an effective vacuum Rabi splitting $g_c$ dictated by the mode volume and the atom-cavity detuning $\Delta_c=2\Delta_{b}$ set by the separation $\Delta_b$ between the atomic frequency and band edge. Thus, exactly as in a real cavity in the far-detuned regime ($|\Delta_c|>|g_c|$) \cite{Agarwal1995}, one atom can exchange its excitation via this virtual `cavity' photon with another atom nearby with a characteristic strength $g_c^2/\Delta_c$. In this case the exchange is characterized by the dipole-dipole interaction Hamiltonian $H=\frac{\hbar g_c^2}{\Delta_c}\sum_{j,l} \sigma^j_{ds} \sigma^l_{sd} f(z_j,z_l)$, where $\ket{s}$ and $\ket{d}$ are the ground and excited states coupled by the cavity mode, see Fig.~\ref{fig:level_int}(c) \cite{John1990a,Kurizki1990a,John1991a,Bay1997a,Lambropoulos2000a,Shahmoon2013a,Shahmoon2014a,Douglas2014a}. 

In the case where the atom is coupled predominantly to only a single band of a one-dimensional photonic crystal waveguide (achieved by tuning the atomic resonance close to the band edge), the interaction has spatial form $f(z_j,z_l) = \exp(-|z_j-z_l|/L)E_{k_b}^*(z_j)E_{k_b}(z_l)$ for atoms at positions $z_j$ and $z_l$, where $E_{k_b}(z)$ is the Bloch function of the photonic crystal at the band edge (wavevector $k_b$, frequency $\omega_b$). Below, we perform numerical simulations with the atoms trapped on regularly spaced lattice with spacing $z_a = 2 a$, for photonic crystal unit cell length $a$, in which case $E_{k_b}(z_j) = 1$ at the trapping sites and the interaction is purely in the relative coordinate $|z_j-z_l|$.
The attenuation length of the interaction $L = \sqrt{\alpha\omega_b/( k_b^2\Delta_b)}$ corresponds to the length of the effective cavity mode and depends on the detuning $\Delta_b$ as well as the curvature of the band edge $\alpha$. The resulting atom-atom interaction can be tuned in strength and length by adjusting these parameters, where a smaller detuning gives a longer range interaction and flatter bands lead to a shorter range interaction.

We wish to convert this dipole exchange interaction into a dispersive interaction between atoms in levels $\ket{s}$, as required to shift the EIT two-photon resonance. To achieve this, we off-resonantly drive the $\ket{s}-\ket{d}$ transition of all atoms with Rabi frequency $\Omega_s$ and detuning $\Delta_s$ ($|\Delta_s|\gg|\Omega_s|$), as shown in Fig. 1(b). In this case, virtual excitations of state $\ket{d}$ result in a Stark shift of level $\ket{s}$ for the $j$th atom, which depends on the number of proximal atoms also in state $\ket{s}$. The interaction now has the general form given by the Hamiltonian $H_{ss} = -\hbar\sum_{j,l} \sigma^j_{ss} \sigma^l_{ss} V(z_j-z_l)$, where $V(z_j-z_l) = -\frac{|\Omega_s|^2g_c^2}{\Delta_s^2\Delta_c}f(z_j,z_l)$ for the single band edge coupling described above. 
For realistic experimental parameters, the coupling $g_c$ can be as large as $2\pi \times$10 GHz for a cavity that is just one wavelength long, $L\sim\lambda$ \cite{Douglas2014a}. Importantly the strength of the interaction can be tuned to any smaller value by altering the ratio $|\Omega_s/\Delta_s|$ for the driving laser, such that level shifts seen by the atoms can be on the order of the freespace linewidth $\gamma$, representing a significant distinction from the Rydberg case.

The photonic crystal mediated interactions are not without losses. In a realistic system imperfections in the dielectric medium cause loss of photons at rate $\kappa$. At the same time the photonic crystals used in current experiments are not three-dimensional and the excited state $\ket{d}$ can spontaneously emit into free space at rate $\Gamma'$ comparable to $\gamma$. To understand the effect of these loss channels we take advantage of the mapping of the photonic cloud surrounding the atom to an effective cavity mode. For an optimal choice of the detuning $\Delta_b$ from the band edge and the curvature $\alpha$ of the band, the two-atom dispersive interaction strength can be made larger than the residual dissipation by a factor $\sqrt{C}/2$, where $C = g_c^2/(\kappa\gamma_d)$ is the cooperativity of the cavity \cite{Douglas2014a}. In state of the art systems the cooperativity can be a large as $C_\lambda \sim 10^4$ for a photonic cloud with an attenuation length $L$ equal to the wavelength $\lambda$.
For longer range couplings with decay length $L$, the cooperativity scales as $C_\lambda\lambda/L$.

\section{EIT with interactions.}\label{sec:eit_interations} 


\begin{figure}
\centering
\includegraphics{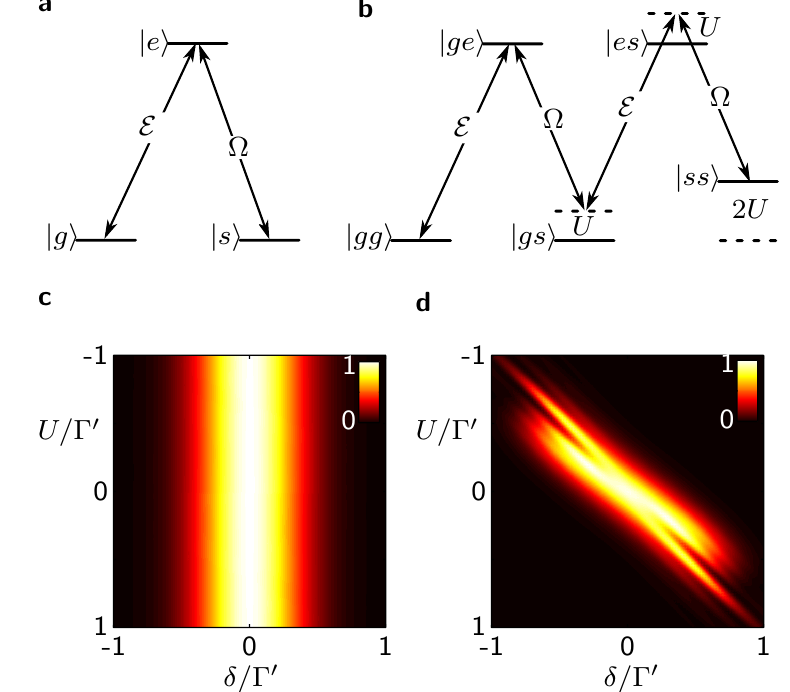}
\caption{EIT in atomic medium with uniform interactions $H_{ss}=\hbar U\sum_{j\neq l} \sigma^j_{ss} \sigma^l_{ss}$ leads to number correlated transparency windows. (a) For a single photon, EIT occurs at the normal two-photon resonance $\delta = 0$. (b) For two photons, EIT occurs at a different two-photon detuning $\delta = U$, in which case transfer to the $\ket{ss}$ two-excitation state is resonant. Simulated normalized (c) output intensity $I_1 = \langle \mathcal{E}_\text{o}^\dagger \mathcal{E}_\text{o} \rangle/\langle \mathcal{E}_\text{i}^\dagger \mathcal{E}_\text{i} \rangle$ and (d) equal-time coincidence rate $I_2 = \langle  \mathcal{E}_\text{o}^\dagger  \mathcal{E}_\text{o}^\dagger  \mathcal{E}_\text{o} \mathcal{E}_\text{o} \rangle/ \langle \mathcal{E}_\text{i}^\dagger \mathcal{E}_\text{i} \rangle^2$ for an EIT system as a function of the strength of the interaction $U$ and the two-photon detuning $\delta$. The system has optical depth $D = 400$ and is continuously pumped with $|\mathcal{E}_\text{i}|^2 = 10^{-4}\Gamma'$ and $\Delta = 0$. (Note that $\delta = U$ can be achieved by adjusting the probe or control field frequencies, here we have adjusted the control as in this case this results in higher two-photon coincidence counts for $|U|>0.5$). For weak input intensity, $I_1$ corresponds to the transmittance of single photons, as the contribution from higher photon number states is negligible. The maximum transmittance occurs when two-photon resonance is satisfied, and is independent of the two-body interaction strength $U$. The coincidence rate $I_2$ shows a maximum when $\delta\sim U$, indicating a maximum transmittance of two photons under this condition. Other simulation parameters used are $\Omega = 2\Gamma'$, $\Gamma_\text{1D}=2\Gamma'$ and $N_a=100$.}
\label{fig:toy_models}
\end{figure}


The basis of EIT in an atomic medium is atoms with three internal levels $\ket{g}, \ket{e}$ and $\ket{s}$, as shown in  Fig.~\ref{fig:level_int}(b). 
We consider a system where the atoms are coupled by the $\ket{g}-\ket{e}$ transition to a quantum probe beam $\mathcal{E}$ propagating in a one-dimensional waveguide, while
the $\ket{e}-\ket{s}$ transition is coupled to an external control laser field (Rabi frequency $\Omega$) and the photonic-crystal mediated interaction acts on level $\ket{s}$ as described above. An experimental system that allows for this configuration has been demonstrated in Refs.~\cite{Goban2013a,Yu2013a,Goban2015a}, where atoms are trapped near a photonic crystal waveguide. In this system the band gaps for the transverse electric (TE) and transverse magnetic (TM) modes of the waveguide occur over different frequency ranges, see Fig.~\ref{fig:level_int}(c). This enables the probe field $\mathcal{E}$ to be guided in the TM mode and couple resonantly to the $\ket{g}-\ket{e}$ transition, while the long-range atom-atom interactions are mediated on the $\ket{s}-\ket{d}$ transition by the TE mode that has a bandgap at the same frequency. At the same time the control beam $\Omega$ may illuminate the atoms from free space. 

With the control laser switched off ($\Omega=0$), and atoms initialized in state $\ket{g}$, probe photons entering the atomic ensemble encounter a two level medium.
When the probe beam frequency $\omega_p$ is on resonance with the atomic transition ($\Delta= \omega_p-\omega_{ge} = 0$), the probe beam transmission is attenuated by a factor of
$\exp(-D)$ after propagating through $N_a$ atoms, where $D = 2 N_a \Gamma_\text{1D}/\Gamma'$ is the optical depth (see Appendix \ref{ap:optical_depth}). Here $\Gamma_\text{1D}$ denotes the emission rate of state $\ket{e}$ into the guided mode of the photonic crystal, while $\Gamma'$ is the emission into all other modes (which in current experiments is comparable to the free-space emission rate $\gamma$) and $\Gamma=\Gamma_\text{1D}+\Gamma'$ is the total single-atom linewidth \cite{Chang2012a}.

Instead when the control is switched on, the two-photon process of absorbing a probe photon and stimulated emission into the control field at frequency $\omega_L$ transfers atomic population to the internal $\ket{s}$ state from state $\ket{g}$. When this two photon process is resonant, $\delta = \omega_p-\omega_L-\omega_{gs} = 0$ (in the absence of atom-atom interactions), interference between excitation from $\ket{g}$ to $\ket{e}$ and $\ket{s}$ to $\ket{e}$ leads to zero population in the excited state and the medium becomes transparent as loss due to spontaneous emission is no longer possible \cite{Fleischhauer2005a}. In this case the input probe photons are mapped to dark state polaritons --- superpositions of electromagnetic fields and spin coherence between the $\ket{g}$ and $\ket{s}$ states \cite{Fleischhauer2000a}. The presence of the atomic spin component allows the speed of propagation to be controlled by the external laser field $\Omega$, where the group velocity becomes $v_g = 2|\Omega|^2/(\Gamma_\text{1D}n)$ for an atomic medium with linear density $n$ in the direction of propagation \cite{Caneva2015a}. For realistic parameters, in the slow-light regime when $v_g\ll c$, the polariton excitations are almost completely spin wave in character, i.e., each polariton excitation effectively results in one atom being flipped to state $\ket{s}$. When atom-atom interactions are not present in the system, this transparency is a linear optical effect and applies for probe pulses of any photon number.

The situation changes dramatically when the photonic crystal interaction $H_{ss}=-\hbar\sum_{j\neq l} \sigma^j_{ss} \sigma^l_{ss} V(z_j-z_l)$ is added (note that we have excluded the diagonal level shift from this Hamiltonian, which is assumed to be absorbed into the definition of our control laser frequency $\omega_L$). In particular, one polariton excitation results in a spatially dependent shift of level $\ket{s}$ for all surrounding atoms. This implies that a second proximal polariton would generally not match the two-photon resonance condition and its propagation is suppressed. Such an effect has already been exploited using Rydberg interactions to produce strong single-photon level nonlinearities \cite{Pritchard2010a,Dudin2012a,Pritchard2013a,Peyronel2012a,Baur2014a,Gorniaczyk2014a,Tiarks2014a}. However, in our case, when the strength and range of interaction associated with $H_{ss}$ are chosen appropriately, a novel phenomenon can appear that has not been previously identified. In particular, it is possible for a second polariton to propagate through, provided that its frequency and shape are specially matched to an altered two-photon resonance condition. In other words, here we will investigate further the possibility of forming correlated transparency windows that depend both on the photon number and wavepacket shape.

We first provide some intuitive examples of such behavior, before considering the physics in greater detail.
We begin with the limit where the interaction is uniform over the atomic sample, which could be achieved by tuning the interaction length $L$ to be much larger than the size of the atomic ensemble so that $V(z_j-z_l)\sim -U$. In this case the second photon sees a constant shift $U$ of the level $\ket{s}$, independent of where the first polariton is \cite{Caneva2015a}. This second photon could satisfy a modified two-photon resonance condition, and thus propagate in a transparent fashion, if its frequency were also compensated by an amount $U$ (for self-consistency, the first photon should also be shifted by frequency $U$ so that both are transparent). More generally, we have a number correlated transparency window, where a single photon propagating by itself will be maximally transmitted through the atomic medium when $\delta = 0$, while $N_p$ photons will be transmitted maximally when $\delta = (N_p-1)U$, see Fig.~\ref{fig:toy_models}(a)-(b).

In Fig.~\ref{fig:toy_models}(c)-(d), we plot measures of single and two-photon output from an EIT medium given an input of continuous coherent light over a range of two-photon detuning $\delta$ and interaction strength $U$. The output probe field $\mathcal{E}_\text{o}$ is produced numerically for a particular input $\mathcal{E}_\text{i}$ by simulating the full EIT system using the spin-model from Ref.~\cite{Caneva2015a}, which we further describe below. We plot the normalized intensity $I_1 = \langle \mathcal{E}_\text{o}^\dagger \mathcal{E}_\text{o} \rangle/ \langle \mathcal{E}_\text{i}^\dagger \mathcal{E}_\text{i} \rangle$, which for weak input probe intensity corresponds to the transmission of single photons through the system. $I_1$ peaks at the normal EIT resonance $\delta = 0$ and is independent of the interaction strength. To visualize the two photon behavior we plot the normalized two-photon coincidence $I_2 = \langle  \mathcal{E}_\text{o}^\dagger  \mathcal{E}_\text{o}^\dagger  \mathcal{E}_\text{o} \mathcal{E}_\text{o} \rangle/ \langle \mathcal{E}_\text{i}^\dagger \mathcal{E}_\text{i} \rangle^2$, which is proportional to the rate at which two photons are detected leaving the system at the same time. $I_2$ peaks when the two-photon transmission is resonant, which occurs when $\delta\sim U$ and confirms the intuition presented earlier. The deviation from unit two-photon transmission for larger values of $U$~($|U|\gtrsim\Gamma'$) is due to the fact that for a continuous wave input, the two photons are entering and leaving at random times. Thus, each photon partially experiences the loss associated with the single-photon dispersion relation at a frequency displaced from the transparency condition \cite{Caneva2015a}.

\begin{figure}
\centering
\includegraphics{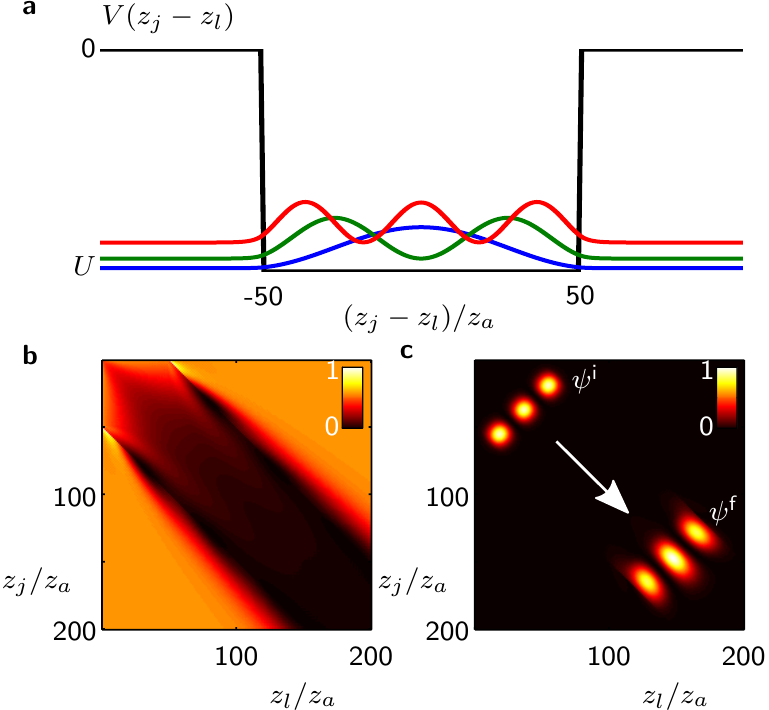}
\caption{EIT in atomic medium with $N_a =200$ atoms on a lattice with lattice constant $z_a$ and interaction $H_{ss}=-\hbar\sum_{j\neq l} \sigma^j_{ss} \sigma^l_{ss} V(z_j-z_l)$. The potential is chosen to be a square well, with an interaction strength of $V(z_j-z_l)=0.5\Gamma'$ for atoms with separation $|z_j-z_l|<50z_a$ and zero otherwise. (a) The three lowest energy states of the step potential. (b) When the system is driven continuously so that $\delta = 0$ ($|\mathcal{E}_\text{i}|^2 = 10^{-4}\Gamma'$), the majority of the steady state spin population $|\psi_{ss}^{j,l}|^2$ (arbitrary units) is in pairs of excitations with $|j-l|>50$. (c) Propagation of a two-photon wave-function  (arbitrary units) in which the relative coordinate is initially prepared as the third harmonic eigenstate (red curve in (a)) of the square-well potential. Here the final spin population is multiplied by a factor of 100, so both initial $\psi^i$ and final $\psi^f$ are visible on the same scale. Other simulation parameters used are $\Omega = 2\Gamma'$, $\Gamma_\text{1D} = \Gamma'$ (optical depth $D=400$) and $\Delta = 0$.}
\label{fig:step_pot}
\end{figure}

An extension of this simple example is to consider a step potential, where the interaction is a constant $U$ up to a certain atom separation $r_s$ and then zero at larger distances, as shown in Fig.~\ref{fig:step_pot}(a). This situation leads to the presence of spatially correlated transparent states of photons, where two photons may propagate together with separation $r < r_s$ for an input such that $\delta \sim U$, or separated with $r > r_s$ when $\delta \sim 0$. 
Since the dark-state polariton excitations essentially consist of atoms excited to state $|s\rangle$, we can visualize this behavior by plotting the probability $|\psi_{ss}^{j,l}|^2$ of having two atoms ($j,l$) both excited to $|s\rangle$. In Fig.~\ref{fig:step_pot}(b) we plot this quantity for the case where $r>r_s$ and $\delta=0$, and in steady state when the system is driven by a weak continuous probe input.
We can then see that when $\delta = 0$ the step potential leads to minimal population of polaritons separated by less than $r_s$. On the other hand when $\delta \sim U$ we expect the polaritons to propagate together within the step separation $r_s$. 
In fact, one expects that if the wave-function of the relative coordinate of a two-photon state is prepared as a harmonic eigenmode of the square well potential (Fig.~\ref{fig:step_pot}(a)), it can propagate maintaining its pulse shape in that direction. This is illustrated in Fig.~\ref{fig:step_pot}(c) for the case where the relative wave-function is initially prepared in the third harmonic. The relative coordinate maintains its shape well, while the wave-function in the center-of-mass coordinate diffuses out.


Motivated by these basic intuitive and numerical observations, we seek to better understand the dynamics of photons within spatial and frequency correlated transparency windows. To this end, a more rigorous description of the EIT system can be developed by finding the effective Hamiltonian for the polariton dynamics in the presence of the atom-atom interactions.  By perturbatively solving the Heisenberg equations of motion around points where such correlated windows are created (as described in Appendix \ref{ap:eff_model}),
we find 
\begin{multline}\label{eq:polariton_H}
H_\text{eff} = \hbar\int dz \Psi^\dagger(z)\left[-i v_g\frac{\partial}{\partial z}-\frac{\hbar}{2 m} \frac{\partial^2}{\partial z^2} 
\right.\\\left.
- \int dz' \Psi^\dagger(z')V_R(z-z')\Psi(z')\right]\Psi(z),
\end{multline}
for the dark-state polariton field-operator $\Psi = \cos\theta \mathcal{E} - \sqrt{n}\sin\theta\sigma_{gs}e^{-i k_p z}$.

This effective Hamiltonian highlights a number of features of polariton evolution, the first being the reduced group velocity, $v_g = c\cos^2\theta\ll c$, of the polariton in the medium that is observed in slow-light experiments, seen here as the first term in the brackets. This term is followed by one describing the dispersion of the group velocity, which is equivalent to the polariton having an effective mass in the EIT medium $m = -\frac{\hbar|\Omega|^2}{(2\Delta_M+ i\Gamma')v_g^2}$. The effective mass is a complex parameter with real part depending on a renormalized probe detuning $\Delta_M$ and the imaginary part on the spontaneous decay rate into other modes $\Gamma'$. The renomalization of the detuning, $\Delta_M = \Delta+\delta$, results from the different bare two-photon detuning $\delta$ required to achieve EIT transparency for different number of photons or spatial configurations in the interacting system. For example, for the case of uniform interactions above, we require $\delta = U(N_p-1)$ for $N_p$ photon transparency and the mass becomes dependent on the number of polaritons in the system \cite{Shahmoon2015a} (see Appendix \ref{ap:eff_model} for further details).  

The final term in the brackets describes the interaction between the polaritons that results from the level-shift caused by the atom-atom interactions. The spin-interaction is also renormalized by the input bare two-photon detuning, where the interaction between the polaritons is described by $V_R(z-z') = \delta + V(z-z')$. It is worth noting that in this case the interaction $ V(z-z')$ does not need to be small or correspond to a level shift within the normal EIT transparency bandwidth (e.g.,~as assumed in Ref.~\cite{Otterbach2013a}). Instead we only require that the renormalized $V_R(z-z')$ fits within the EIT transparency within the spatial region where there is significant polariton population. Indeed Eq.~(\ref{eq:polariton_H})  can describe different self-consistent transparency windows, for different photon number and spatial configurations, and these need not overlap.

For two photons, Eq.~(\ref{eq:polariton_H}) separates into center of mass and relative coordinates, where each coordinate has a mass/diffusion term exactly like in non-interacting EIT. In particular, a well-localized wave-function experiences a similar loss as in normal EIT. However, the relative coordinate also sees the effective potential, which could, for example, support bound states as we describe in the next section. In general we can map a multi-polariton system to a collection of massive particles in a potential moving in a reference frame traveling at $v_g$.

The effective Hamiltonian of Eq.~\ref{eq:polariton_H} is valid around the two-photon resonance where we expect it to provide an approximate description of the full system dynamics. To confirm these dynamics we numerically simulate the full system using the spin model introduced in Ref.~\cite{Caneva2015a}. The spin model describes the behaviour of many atoms coupled to the photonic modes of a one-dimensional waveguide, generalizing the powerful input-output formalism of cavity QED. The key insight of the model is that the exchange of photons (wavevector $k_p$) between atoms via the waveguide reduces to an effective interaction between the atoms, given by 
\begin{equation}
H_\text{wg} = - \frac{i\hbar\Gamma_\text{1D}}{2}\sum_{jl}\sigma^j_{eg}\sigma^l_{ge}e^{i k_p|z_i-z_j|}.
\label{eq:spin_model}
\end{equation}
The dynamics of the electromagnetic field as it propagates through the system are now completely captured in the dynamics of the interacting spin chain with Hamiltonian $H_\text{spin} = H_\text{wg} + H_\text{atom} + H_\text{pump} +  H_{ss}$, where $H_\text{atom}$ describes the atomic Hamiltonian in the absence of the waveguide mediated interactions and $H_\text{pump}$ describes the coupling of the atoms to input light. 
Below we consider the case where the input probe light is coherent, in which case the coupling to the atoms can be accounted for by using the Mollow transformation \cite{Mol75}, and we have $\mathcal{H}_\text{pump}=  -\hbar\mathcal{E}_\text{i}(t) \sqrt{\frac{\Gamma_{1d}}{2}}\sum_j (\sigma^j_{eg} e^{i k_p z_j} +  \sigma^j_{ge} e^{-i k_p z_j})$.
Here $\mathcal{E}_\text{i}(t)$ is a classical field that may either be constant, corresponding to continuous driving of the system, pulsed or zero when we initialize the system with a spin wave. 


\begin{figure}
\centering
\includegraphics[scale=0.95]{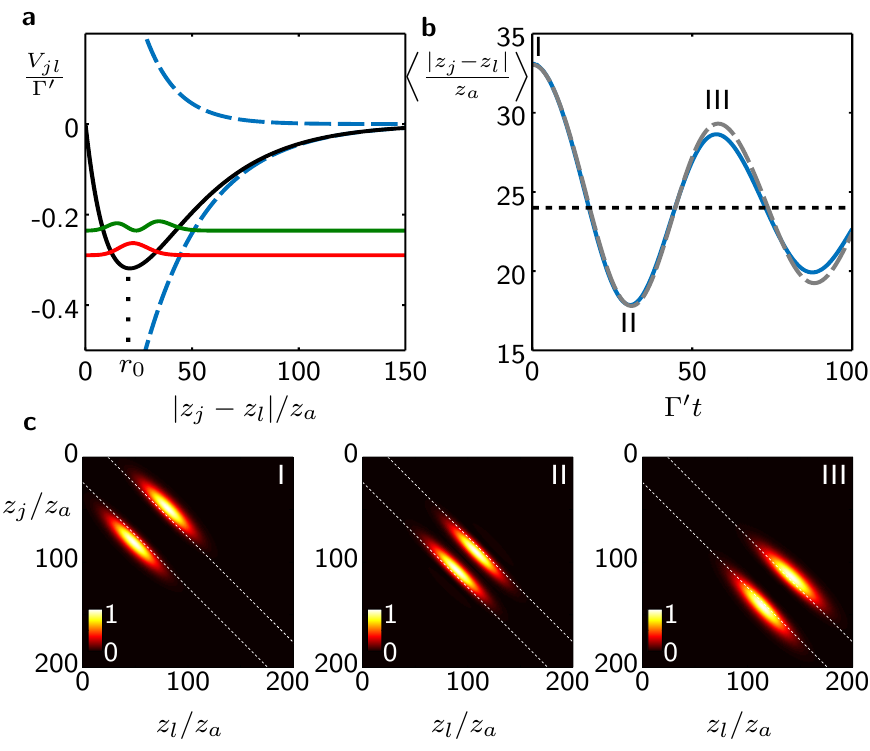}
\caption{(a) The photonic crystal mediated atomic interactions result in a interaction potential $V_{jl} = V(z_j-z_l) = G[\exp(-|z_j-z_l|/L_u) - \exp(-|z_j-z_l|/L_\ell)]$ (black solid line) for polaritons separated by distance $|z_j-z_l|$. The interaction potential is the sum of the contribution from the lower and upper photonic crystal band edges (dashed blue lines) that have length scales $L_u = 15 z_a$ and $L_\ell = 30 z_a$ and stength $G = 1.28\Gamma'$, which in this case leads to a minimum at a finite atomic separation $r_0 \sim 21 z_a$. Loss resulting from the photonic crystal mediated interactions is calculated assuming a cooperativity of $C_\lambda = 24000$. The amplitudes $|\psi(r)|^2$ (arbitrary units) of the two lowest bound-state wavefunctions are shown for $\Delta_M = 2.5\Gamma'$, $\Omega =\Gamma'$, and $\Gamma_\text{1D} = 2\Gamma'$. (b) For two polaritons propagating with this interaction, a state initiated with wavefunction matching that of the ground state, but offset from the ground state separation at time $t=0$ oscillates in time in the relative coordinate. The oscillation of the expectation value of polariton separation about the ground state separation ($\sim 24z_a$ indicated by the dotted line) found by simulation the full EIT system (solid line) is closely matched to the oscillation produced by the evolution of the effective Hamiltonian $H_\text{rel}= -\frac{\hbar^2}{m}\frac{\partial^2}{\partial r^2}- 2\hbar V(r)$ (dashed line). In this case a full oscillation occurs after the polaritons travel an optical depth of $D = 250$ using the parameters in (a). (c) Spin population $|\psi^{ss}_{j,l}|^2$ (arbitrary units) as the initially offset pair of polaritons propagate through the atomic ensemble, with the dotted line indicating the ground state separation of the two polaritons. The frames I,II,III correspond to the turning points of the oscillation marked in (b).}
\label{fig:potential}
\end{figure}


The output field operator 
$\mathcal{E}_\text{o}(z,t) = \mathcal{E}_\text{i}(t-z/c) + \varepsilon_\text{i}(t-z/c) + i\sqrt{\frac{\Gamma_{1d}}{2}}\sum_{j=1}^N\sigma^j_{ge}(t)e^{i k_p(z-z_i)}$
is determined from the coherent input, with classical part $\mathcal{E}_\text{i}$ and vacuum fluctations $\varepsilon_\text{i}$, and the atomic coherence $\sigma^j_{ge}$ found by solving the spin evolution.
The numerical description of the system is then reduced to following the dynamics of the discrete atomic Hilbert space under $H_\text{spin}$ instead of the in-principle continuous electromagnetic field. This allows us to fully model our EIT system including arbitrary interactions between the atoms in a way that is independent of the assumptions made in deriving the effective polariton dynamics in Eq.~(\ref{eq:polariton_H}), providing numerical verification of the intuitive picture of polaritons as massive interacting particles.

\section{Photonic molecules}\label{sec:photon_molecules}

We now use our effective propagation equation along with the spin model to describe a regime where photons can bind to one another in the EIT medium. Here we rely on the tunability of the interactions in the photonic crystal setting to achieve a molecular-like interaction potential between polaritons. In particular, we consider the case where our photonic crystal has two band edges close to one another (see Fig.~\ref{fig:level_int}(b)) to which the $\ket{s}-\ket{d}$ transition couples. In this case the resulting effective interaction  $V(z_j-z_l) = -\frac{|\Omega_s|^2}{\Delta_s^2}\left[\frac{g_u^2}{\Delta_u}\exp(-|z_j-z_l|/L_u) + \frac{g_\ell^2}{\Delta_\ell}\exp(-|z_j-z_l|/L_\ell)\right]$ will be the sum of the interaction due the upper ($u$) and lower  ($\ell$) band edges (see Appendix \ref{ap:potential}). Crucially, the sign of each contribution depends on the detuning from the band edge in question, and the contributions hence have opposite signs. Furthermore, by engineering the detunings and band curvatures correctly the contributions will also have different interaction length scales, leading the total interaction to have a minimum (or maximum) at some atomic separation $r_0$, as plotted in Fig.~\ref{fig:potential}(a) for the case where the interaction strengths have equal magnitude $G \equiv \frac{|\Omega_s|^2}{\Delta_s^2}\frac{g_u^2}{\Delta_u} = \frac{|\Omega_s|^2}{\Delta_s^2}\frac{g_\ell^2}{\Delta_\ell}$.   

For the case of two polaritons, with coordinates $z_1$ and $z_2$, Eq.~\ref{eq:polariton_H} separates into a Hamiltonian $H_\text{cm}= -\frac{\hbar^2}{4m}\frac{\partial^2}{\partial R^2}- i\hbar v_g \frac{\partial}{\partial R}$ for the center-of-mass coordinate $R = (z_1+z_2)/2$, whose dispersion corresponds to that of a free massive particle, and a Hamiltonian $H_\text{rel}= -\frac{\hbar^2}{m}\frac{\partial^2}{\partial r^2}- 2\hbar V(r)$ for the relative coordinate $r = z_1-z_2$. From this we can find the bound states of pairs of polaritons in the system, as shown in Fig.~\ref{fig:potential}(a). Initializing the two-polariton wavefunction in the motional ground state of the potential then leads to the two polaritons propagating through the system with fixed relative position.  We may also perturb the state by off-setting the initial relative position, in which case we observe phononic molecular vibrations about the ground state separation (slightly offset from $r_0$ due to the asymmetry of the potential). In Fig.~\ref{fig:potential}(b) we plot the decaying oscillation of the relative position about the ground state separation for a full spin model simulation of a two polariton wavefunction that has an initial offset and see the good agreement with the oscillation predicted by the simple model. In Fig.~\ref{fig:potential}(c) the polariton wavepacket in the two excitation manifold is plotted at the extrema of the oscillations, where we see that the relative coordinate remains tightly bound, while the wave function in the center-of-mass coordinate disperses as a free massive particle. 

\begin{figure}
\centering
\includegraphics{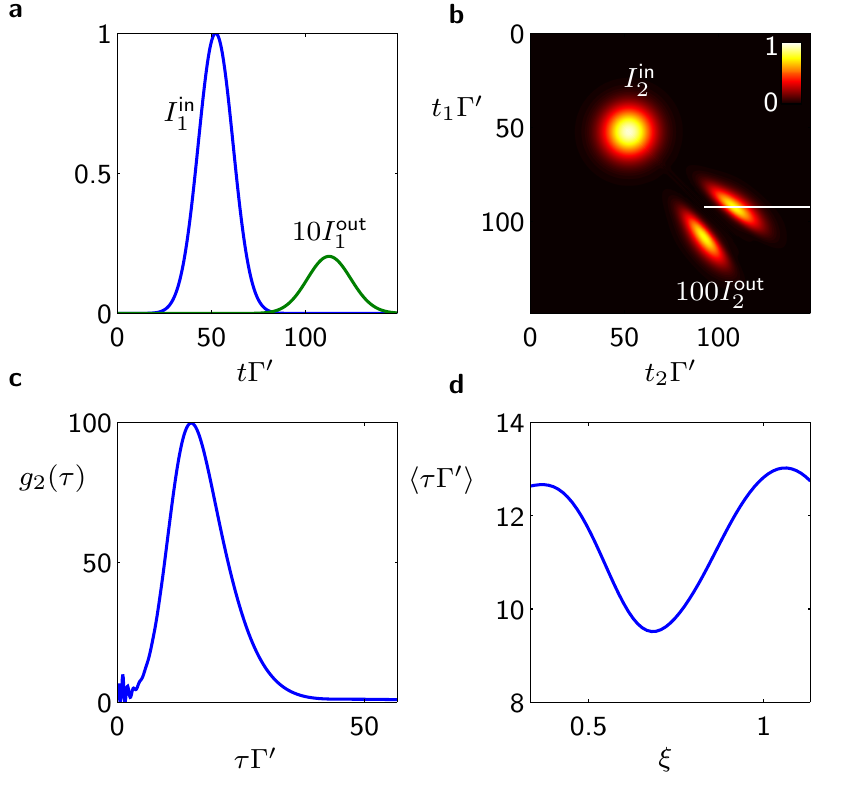}
\caption{Input and output of an EIT system with molecular-like interactions resulting from the potential shown in Fig.~\ref{fig:potential}(a). (a) Single photon component of input (blue) and delayed output (green) pulses, normalized by the peak intensity of the input pulse, where the output has been multiplied by a factor of 10 for visibility. (b) Two-photon input and output components where the output has been scaled by a factor of $10^2$. (c) Second order correlation function $g_2(\tau)$ for output two-photon separation time $\tau = |t_1-t_2|$ taken along the white line shown in (b). (d) Oscillation in average two-photon separation time $\langle\tau\rangle$ at output as the interaction strength and renormalized detuning are scaled to become $G' = \xi G$ and  $\Delta_M' = \xi \Delta_M$. The optical depth is 400 for plots (a)-(c) and 800 for plot (d), with $G = 1.28\Gamma'$, $\Delta_M = 1.5\Gamma'$, $\Omega =1.5\Gamma'$, $|\mathcal{E}_\text{i}|^2 = 10^{-4}\Gamma'$ and $\Gamma_\text{1D}= 2\Gamma'$.}
\label{fig:full_pot}
\end{figure}

In an experiment this type of oscillation may be observable by spin-dependent imaging of the cold atoms; however, the most straight forward measurements are of photons output from the system into the waveguide. Using the spin model, we can simulate an experiment where a weak coherent pulse is input into the system at a frequency resonant with the two photon molecule ($\delta \sim V(r_0)$) and the output fields are recorded after the photons are allowed to propagate through the system. In Fig.~\ref{fig:full_pot}(a)-(b) we plot the single and two-photon parts of the input and output fields for a weak pulse with a Gaussian spatial envelop. Here, we see that the correlated transparency window damps most of the two-photon component of the initial Gaussian input, and what remains at the output largely coincides with the well-separated two-photon molecule. This leads to a peak in the second order correlation function $g_2(\tau) = \langle \mathcal{E}_\text{o}^\dagger(t)\mathcal{E}_\text{o}^\dagger(t+\tau)\mathcal{E}_\text{o}(t+\tau)\mathcal{E}_\text{o}(t)\rangle/(\langle \mathcal{E}_\text{o}^\dagger(t)\mathcal{E}_\text{o}(t)\rangle\langle \mathcal{E}_\text{o}^\dagger(t+\tau)\mathcal{E}_\text{o}(t+\tau)\rangle)$ for the photons, Fig.~\ref{fig:full_pot}(c), which shows large bunching due to the presence of the bound photon pair.


We can also observe the phononic oscillations of the pair of photons in the photon output. Inputting photons into the system in a Gaussian pulse as above, the photons are initially separated on average by less than the ground state separation, which initializes an oscillation in the two-photon wavefunction. The relative position of the two photons at the output depends on the ratio of the length over which the pulse travels in one oscillation period, $L_o$, to the length of the atomic medium $L_s$. 
An oscillation can then be observed at output by either adjusting the effective length of the system, for example by reducing the number of atoms by waiting for the atomic trap to decay, or by adjusting the strength $G$ of the interaction potential, for example by adjusting the drive Rabi frequency $\Omega_s$. In the latter case, the oscillation length is proportional to $1/\sqrt{\Delta_M G}$. However, adjusting the strength $G$ alone also changes the shape of the molecule, as the spatial extent of the ground state is approximately proportional to $\sqrt{\Delta_M/G}$. Reducing both $\Delta_M$ and $G$ by the same factor $\xi$ then results in the molecule keeping its shape while increasing $L_\text{o}$ by a factor of $1/\xi$. In Fig.~\ref{fig:full_pot}(d) we show a simulation of this situation, where we adjust $G$ and $\Delta_M$ from the values used in Fig.~\ref{fig:full_pot}(a)-(c) that initially lead to approximately one and a half oscillations over the length of the system. By reducing the values by a factor of three the exact simulation shows a full oscillation in the two photon output, i.e., $L_\text{o}$ changes from $3/2L_s$ to $1/2L_s$, confirming the prediction of our simple model.

\section{Technical requirements}\label{sec:tech_req}

Having described the main features of the two-photon molecule in the previous section, we now analyze more carefully the technical requirements in order to observe this physics. There are two main considerations, the first being the regime of validity of the effective equation (\ref{eq:polariton_H}) that describes polaritons as massive particles propagating in a potential, and the second, the extent to which dissipation affects the molecular state. Describing the polariton dynamics by Eq.~(\ref{eq:polariton_H}) is equivalent to keeping only terms of up to quadratic order in an expansion of the photon dispersion relation about the EIT resonance. Neglecting higher order terms is a good approximation within the EIT window, which for $|\Delta| > |\Omega|, \Gamma'$, requires that the spread of frequencies of the relevant system dynamics remains within $|\Omega|^2/|\Delta|$ of the EIT resonance. For a pulse that has length $z_0$ within the atomic medium (e.g., a Gaussian intensity distribution $I(z) \propto \exp(-8 z^2/z_0^2)$) to be described by Eq.~(\ref{eq:polariton_H}) it must have frequency width $d\omega \sim 4 v_g/z_0 < |\Omega|^2/|\Delta|$, which can be recast as a requirement $D_p>16 |\Delta|/\Gamma'$ for the amount of optical depth $D_p$ in the pulse length. This identifies a tradeoff, where we would like to increase the ratio $|\Delta|/\Gamma'$ to reduce the effect of dissipation in the effective mass term, however to do so requires an increase in the optical depth required in an experiment.

Dissipation in our system comes from two main sources, one being spontaneous emission from the excited state $\ket{e}$ that gives rise to the imaginary part of the effective mass term, and the other being the additional loss associated with introducing the photonic crystal mediated interactions. The former leads to loss as a pulse propagates over length $L$ in the atomic medium, which in the limit of validity of the quadratic dispersion is approximately $\exp[-16 L/(z_0 D_p)]$. This is the loss for linear propagation of a pulse and for two photons propagating in a molecule the loss is squared. To observe an oscillation of a photonic molecule we would then like to take a minimal value of the oscillation length $L_o$ to reduce loss. To see how $L_o$ is constrained, we consider the case where the interaction potential can be approximately described as a harmonic oscillator, with Gaussian ground state wavefunction $\exp[-2 (r-r_0)^2/(z_0^2)]$ in relative coordinate space, in which case the photon molecule would oscillate with frequency $\omega_M  \approx \frac{32 v_g|\Delta|}{\Gamma_\text{1D} n z_0^2}$. The oscillation length $L_o  = 2 \pi v_g/\omega_M$ is reduced by increasing $\omega_M$, however this frequency is constrained to remain within the EIT window, which then results in the restriction $L_o > \pi z_0/2$. Taking the minimum value for $L_o$ and equating this with the system length $L_s$, such that the photon molecule undergoes one full oscillation, 
then yields a propagation loss of $\exp[-16 \pi /D_p]$.

As noted in Section \ref{sec:phc_int} the loss related with the photonic crystal mediated interactions depends on the cooperativity of the atom-induced cavities in the photonic crystal. For a potential like that in Fig.~\ref{fig:potential}(a), the loss rate is approximately $\beta\omega_M\sqrt{z_0/(\lambda C_\lambda)}$ (taking $r_0\sim z_0$) where the proportionality constant $\beta$ depends on the details of how the two potentials are combined (see Appendix \ref{ap:potential}). Over one oscillation period of the photonic molecule we then expect the norm of the two photon wavefunction to be reduced by a factor of approximately $\exp[-\frac{2 \pi\beta\sqrt{z_0}}{\sqrt{\lambda C_\lambda}}]$ as a result of this loss. Compared with the propagation loss, this loss increases with increasing pulse length $z_0$, as larger constituent pulses require larger effective cavities to create the potential in the photonic crystal. 
The total loss then represents a balance between the two contributions and an optimal pulse length can be found $z_0 = 2(4C_\lambda\Gamma'^2/(\beta\Gamma_\text{1D})^2)^{1/3}$, at which point the total loss is $\exp(-6 \pi (2\beta ^2 \Gamma'/(C_\lambda\Gamma_\text{1D}))^{1/3})=\exp(-48\pi/D_p)$. For parameters compatible with current experimental setups, $\Gamma_\text{1D} = \Gamma'$, $\beta\sim 10$ and $C_\lambda = 2\times 10^4$, we have $z_0 \sim 20a$ and the $\exp(-48\pi/D_p) \sim 0.02$.
The total optical depth to observe a single oscillation with these parameters is then $\sim 140$. In experimental systems to date only a few atoms have been trapped near the photonic crystals, however optical depths per atom of $D_a \sim 2$ have already been demonstrated \cite{Goban2015a}, indicating that sufficient optical depth would be available with $\sim 70$ atoms.

While we have focused on the loss of the photonic molecule above, signatures of the molecule and its oscillations can be observed in photon correlation measurements with greatly reduced requirements on the system parameters. In particular, for a weak coherent state input, the normalized second-order correlation function $g^{(2)}(\tau)$ will contain the salient features of the molecular dynamics even with significant two-photon losses, provided that the single-photon component decays in an even stronger manner. This strong single-photon loss can occur by exploiting the design of the correlated transparency window, and in particular enforcing that the transparency frequency for the two-photon molecule coincides with a region of the single-photon spectrum exhibiting strong absorption.  This results in large loss of the single photon component, leading to the large values of $g_2(\tau)$ shown in Fig.~\ref{fig:full_pot}(c), where our simulations include the loss mechanisms described above. 

The behavior of $g_2(\tau)$ also depends on the nature of input and output. In particular, since the two-photon molecule is comprised of spatially separate photon pulses, this implies that there is a time $\sim r_0/v_g$ upon entering and exiting that only one of the photons is inside the medium. During this time, the two-photon component experiences loss as a single photon would.  This additional loss to the two-photon wavepacket could be avoided if the pulses were first loaded into the medium and then the interactions switched when both photons had already entered the medium and a similar procedure applied at output.

\section{Summary and conclusions}\label{sec:summary}

In summary we have demonstrated that the versatile platform of cold atoms coupled to photonic crystal waveguides leads to new possibilities in quantum optics. In particular, the ability to tune interactions between atoms over distances much greater than the wavelength of light allows the propagation of light through the atomic medium to be highly non-linear and non-local. In contrast with current experiments demonstrating non-linearities due to Rydberg interactions between atoms \cite{Pritchard2010a,Dudin2012a,Pritchard2013a,Peyronel2012a,Baur2014a,Gorniaczyk2014a,Tiarks2014a}, the interactions discussed here can be arranged to give level shifts of the order of the atomic linewidth over the entire atomic ensemble. This allows for correlated transparency of photons through the medium, where depending on the number of photons propagating in the system, the system is only transparent to particular frequencies and spatial configurations, for example a photon molecule. While the focus of our numerical studies here have mainly been in comparing single and two photon dynamics in the system, the intuition formed provides important insight into the solution of the multi-photon problem, which is expected to display rich many-body behavior \cite{Otterbach2013a,Shahmoon2015a}.

\section{Acknowledgements}

The authors thank E. Shahmoon, A. Gorshkov and the Kimble group for stimulating discussions. This work was supported by  Fundacio Privada Cellex Barcelona, the MINECO Ramon y Cajal Program, the Marie Curie Career Integration Grant ATOMNANO, and ERC Starting Grant FoQAL.

\appendix

\section{Optical depth}\label{ap:optical_depth}

In the main text we use the form $D = 2 N_a \Gamma_\text{1D}/\Gamma'$ to denote the resonant optical depth for $N_a$ two-level atoms.
This expression can be derived using the transfer matrix formalism, where linear propagation in one-dimension is well described by treating each atom as a refractive element with transmission and reflection coefficients $t = 1 + r$ and $r = -\Gamma_\text{1D}/(\Gamma_\text{1D}+\Gamma')$ on resonance \cite{Chang2011a,Chang2012a}. For $\Gamma_\text{1D}\ll \Gamma'$ one expects that reflection is not important, and the transmitted intensity is given by the product of the individual atomic transmittance as $\sim |t|^{2 N_a} \sim \exp(-D)$. However when $\Gamma_\text{1D} \sim \Gamma'$ a single atom can act as a strong reflector of light and the spacing of the atoms becomes crucial in how much light is transmitted through the sample \cite{Chang2012a,Chang2007a}. In our simulations we choose the spacing between the atoms $z_a$ so that $k_0 z_a = 3\pi/2$, in which case there is destructive interference between the reflections from each atom in the ensemble. The transmission is then reproduced by the expression $\exp(-D)$ for $\Gamma_\text{1D} \lesssim 0.2 \Gamma'$. Under this condition the optical properties are the same for two samples with the same $D = 2 N_a \Gamma_\text{1D}/\Gamma'$, and using this property we can may then reduce the number of atoms required to simulate a system with particular optical depth by increasing $\Gamma_\text{1D}$ proportionally.

The linear properties in the EIT system can be even more robust, remaining the same for constant optical depth even when $\Gamma_\text{1D} \sim \Gamma'$. A single three-level atom has transmission and reflection coefficients $t = 1 + r$ and $r = -\Gamma_\text{1D}\delta/[(\Gamma_\text{1D}+\Gamma'-2 i \delta)\delta + 2 i \Omega^2]$ for $\Delta =0$. We then find that after propagating through $N_a$ atoms with spacing $k_0 z_a = 3\pi/2$ the probe field acquires a phase given by $i\exp[-N_a \Gamma_\text{1D}\delta/[(\Gamma'-2 i \delta)\delta + 2 i \Omega^2]$, provided the two-photon detuning $\delta$ satisfies the condition $|\delta| \ll 2|\Omega|^2/\Gamma_\text{1D}$. As in the two level case above this phase factor does not depend on $\Gamma_\text{1D}$ and $N_a$ individually but on their product. Thus for probes with $|\delta| \ll 2|\Omega|^2/\Gamma_\text{1D}$ we can again reduce the number of atoms and increase $\Gamma_\text{1D}$ in a simulation to reduce computational requirements. From a practical point of view we can check this equivalence in our simulations by changing $\Gamma_\text{1D}$ and $N_a$ appropriately and observing that the simulated behavior does not change.

\section{Effective propagation of polaritons}\label{ap:eff_model}

When photons enter the EIT medium they are converted into polaritons, where the photonic excitation becomes a mixture of a photonic part and spin wave part with atomic population in the level $\ket{s}$.
A single photon entering the system sees the EIT resonance when $\delta = 0$ (see Fig.~\ref{fig:level_int}b). For more than one photon, the presence of the interaction between the $\ket{s}$ levels shifts the EIT resonance. A shift in the EIT resonance will lead to changes in velocity and dispersion of each pulse, and with a spatially dependent potential as shown in Fig.~\ref{fig:potential}a these effects will depend on the separation of polaritons in the media. However, it is not immediately clear from the Hamiltonian how and if the atom-atom interaction potential transforms into an interaction potential for polaritons. To understand more about the system we now derive an effective propagation equation for polaritons in the system.

To find the effective equation, we model the effective three-level media in the continuum limit, with linear atomic density $n$, coupled to the probe with slowly varying envelop $\mathcal{E}(t,z)$. In the frame rotating with the input frequency the system is described by the Hamiltonian \cite{Hafezi2012a}
\begin{multline}\label{eq:H_three_lev}
 \mathcal{H}= -n \hbar\int dz \left\{\left(\Delta + i\frac{\Gamma'}{2}\right)\sigma_{ee}(z) +  
 \delta\sigma_{ss}(z) + \right.\\\left.
\left[\sigma_{es}(z)\Omega e^{i k_c z} + \sqrt{\frac{c\Gamma_\text{1D}}{2}}\sigma_{eg}(z)\mathcal{E}(z) e^{i k_p z}
+ \mbox{H.c.}\right]\right\} \\-n^2\hbar\int dz dz' \sigma_{ss}(z)V(z,z')\sigma_{ss}(z').
\end{multline}
The Heisenberg equations of motion for the probe field and the atom coherences are then, to lowest order in the probe field, 
\begin{align}
&\frac{\partial\tilde{\sigma}_{ge}(z)}{\partial t} = i\left(\Delta + i\frac{\Gamma'}{2}\right)\tilde{\sigma}_{ge}(z) + i \sqrt{\frac{c\Gamma_\text{1D}}{2}} \mathcal{E}(z)+ i \Omega\tilde{\sigma}_{gs}(z)\nonumber\\
&\frac{\partial\tilde{\sigma}_{gs}(z)}{\partial t} = i\nu(z)\tilde{\sigma}_{gs}(z) + i \Omega^*\tilde{\sigma}_{ge}(z)\nonumber\\
&\frac{\partial\mathcal{E}(z)}{\partial t} + c\frac{\partial\mathcal{E}(z)}{\partial z}= i\sqrt{\frac{c\Gamma_\text{1D}}{2}} n \tilde{\sigma}_{ge}(z)
\end{align}
where $\nu(z) = \delta+2n\int dz' \sigma_{ss}(z')V(z,z')$, $\tilde{\sigma}_{ge}(z) = \sigma_{ge}(z)e^{-i k_p z}$ and $\tilde{\sigma}_{gs}(z) = \sigma_{gs}(z)e^{i (k_c-k_p) z}$.

We now introduce the dark and bright-state polariton operators, $\Psi = \cos\theta \mathcal{E} - \sqrt{n}\sin\theta\tilde{\sigma}_{gs}$ and $\Phi =  \sin\theta\mathcal{E} +  \sqrt{n}\cos\theta\tilde{\sigma}_{gs}$ \cite{Otterbach2013a,Fleischhauer2002}. The mixing angle is given by $\tan\theta = \sqrt{\frac{c n \Gamma_\text{1D}}{2|\Omega|^2}}$ ($\approx \sqrt{c/v_g}$ in the slow light regime). We then find the following equations for the polariton operators
\begin{multline}
\frac{\partial\Psi}{\partial t} + c\cos^2\theta\frac{\partial\Psi}{\partial z} =  i\sin^2\theta\nu(z)\Psi \\
-\cos\theta\sin\theta\left[c\frac{\partial\Phi}{\partial z}+ i\nu(z)\Phi\right]
\end{multline}
and
\begin{multline}
\Phi =  \frac{\cos^2\theta}{|\Omega|^2}\left(i\frac{\partial}{\partial t}+\nu(z)\right)\left(i\frac{\partial}{\partial t}+\Delta+i\frac{\Gamma'}{2}\right)
 \\\times(\Phi-\tan\theta\Psi)
\end{multline}
with $\nu(z) \approx \delta + 2\int dz' \Psi^\dagger(z')V(z-z')\Psi(z')$.

In the slow light limit where $v_g = c \cos^2\theta\ll c$ and $\sin^2\theta\sim 1$, the bright state polariton has a pertubative effect on the propagation of the dark state polariton \cite{Fleischhauer2002}. At lowest order in $\sqrt{v_g/c}$ we get the effective propagation equation for the dark state
\begin{multline}
\left(\frac{\partial}{\partial t} + v_g\frac{\partial}{\partial z}+i\frac{v_g^2}{|\Omega|^2}\frac{\partial^2}{\partial z^2}\left(\Delta-\nu(z) + i\frac{\Gamma'}{2}\right)\right.\\\left.+\frac{v_g^3}{|\Omega|^2}\frac{\partial^3}{\partial z^3}\right)\Psi=
 i \nu(z)\Psi.
\end{multline}
We can make further approximations, neglecting the third spatial derivative term in the limit where the second spatial derivative terms dominate, that is when $|\Delta+i\Gamma'/2|\gg v_g/z_0$ for a polariton with pulse length $z_0$. The $\nu(z)$ part of the second spatial derivative term can also be simplified in certain limits leading to an effective mass. Particularly, in the case of uniform interactions there is no spatial dependence and it may be taken outside of the derivative. In the photon molecule case discussed in the main text, the spatial dependence is approximately $\omega_p z^2/z_0^2$, which leads to terms that scale with $\omega_p/D_p^2$ that can be neglected for large optical depth pulses. In this case we may replace $\nu(z)$ by $\nu(r_0)$. 

The exact form of the effective mass then depends on the particular correlated EIT transparency we operate around, and is typically number dependent \cite{Shahmoon2015a}. For example, when we have uniform interactions $\nu(z) = \delta - 2 U (N_p-1)$ and when we are expanding around the transparency for $N_p$ photons, that is when $\delta = U(N_p-1)$, we have that the effective mass term $m = -\frac{\hbar|\Omega|^2}{(2\Delta_M + i\Gamma')v_g^2}$ depends on the renormalized detuning $\Delta_M = \Delta+U(N_p-1)$ (see for example Fig.~\ref{fig:toy_models}(b)). Similarly in the photon molecule case we have $\nu(z) \approx \nu(r_0) = \delta + 2 V(r_0)$ and driving at the molecule transparency, i.e., $\delta \sim -V(r_0)$, leads to the effective mass depending on $\Delta_M = \Delta-V(r_0)$. In both cases we have the renormalized detuning $\Delta_M = \Delta+\delta$, and we arrive at the Hamiltonian in Eq.~(\ref{eq:polariton_H}).

\section{Inter-atomic potential}\label{ap:potential}

As described in the main text the interaction of atoms via a single band edge leads to the effective interaction Hamiltonian 
$H = \frac{\hbar|\Omega_s|^2g_c^2}{\Delta_s^2\Delta_c}\sum_{j,l} \sigma^j_{ss} \sigma^l_{ss} \exp(-|z_j-z_l|/L)E_{k_b}^*(z_j)E_{k_b}(z_l)$. For structures like the ``alligator'' photonic crystal waveguide described in Ref.~\cite{Yu2013a}, the Bloch functions near the band edge are well approximated by the form $E_{k_b}(z)\sim \cos(k_b z)$, in which case if the atoms are trapped at sites commensurate with the band-edge wavelength we have $E_{k_b}(z)\sim 1$.

If the atoms now interact via two distinct band edges, as in Fig.~\ref{fig:level_int}(b), the interaction potential $V(z_j-z_l)$ becomes the sum of the contributions from the upper ($u$) and lower ($\ell$) band edges, $V(z_j-z_l) = -\frac{|\Omega_s|^2}{\Delta_s^2}\left[ \frac{g_u^2}{\Delta_u} \exp(-|z_j-z_l|/L_u) + \frac{g_\ell^2}{\Delta_\ell} \exp(-|z_j-z_l|/L_\ell)\right]$. The two contributions are each related with effective cavity parameters $g_{u,\ell}$, $\Delta_{u,\ell}$ and $L_{u,\ell} = \sqrt{2\alpha_{u,\ell}\omega_b/( k_b^2\Delta_{u,\ell})}$. By adjusting the detunings $\Delta_{u,\ell}$ and band curvatures $\alpha_{u,\ell}$ we can tune the potential to have a minimum at some finite particle separation $r_0$, as in Fig.~\ref{fig:potential}(a).

We can now consider the loss introduced into our system for a single atom in the $\ket{s}$ state as a by-product of creating the inter-atomic potential above. As discussed in the main text (and in more detail in Ref.~\cite{Douglas2014a}) the loss due to each band edge is the same as the loss that would result if the interaction were created by a cavity with the same parameters. In particular in the detuned regime $|\Delta_{u,\ell}| > |g_{u,\ell}|)$, the loss from spontaneous emission $\gamma_d$ and cavity decay $\kappa$ occurs at rate approximately $\frac{|\Omega_s^2|}{\Delta_s^2}\left(\gamma_d + \kappa \frac{g_{u,\ell}^2}{\Delta_{u,\ell}^2}\right)$. This loss can then be minimized with respect to the interaction strength $G_{u,\ell} = \frac{|\Omega_s|^2}{\Delta_s^2} \frac{g_{u,\ell}^2}{\Delta_{u,\ell}}$ by adjusting the detuning, in which case the loss rate from each cavity for an atom in the $\ket{s}$ state becomes $2G_{u,\ell}/\sqrt{C_{u,\ell}}$ for cavity cooperativity $C_{u,\ell} = g_{u,\ell}^2/(\kappa\gamma_d)$.

We can further compare the rate of loss due to these effective cavities to the energy of the two photon bound state, which gives the rate of oscillation of photon molecule. The relation between the strengths of the individual potentials due to each band edge and the molecule energy $\omega_M$ depends on how the potentials add together and will depend on the desired shape and experimental constraints. As an example we consider the case where the band edges are arranged such that $G_u=G_\ell\equiv G$, in which case the contribution from the potential to the dressed state energy of individual atoms is zero. The potential is now $V(z_j-z_l) = G[\exp(-|z_j-z_l|/L_u) - \exp(-|z_j-z_l|/L_\ell)]$ and choosing for example $L_\ell = 8 L_u$ a potential minimum occurs at $r_0\sim 0.3 L_\ell$ with depth $\sim 2 G/3$, which provides an upper limit on the energy of the two-photon molecule energy $\omega_M$. Optimizing the detunings as above we then find that the loss is $\sim \beta \omega_M\sqrt{r_0/(\lambda C_\lambda)}$ with $\beta = 10$.

The potential we show in Fig.~\ref{fig:potential} has $L_u = 15 a$ and $L_\ell = 30 a$, where using a smaller ratio between $L_\ell$ and $L_u$ puts less demands on the engineering of band curvatures, since for $G_u = G_\ell$ we have that $L_\ell/L_u = \alpha_\ell/\alpha_u$. The strength of the potential is $G = \frac{|\Omega_s|^2}{\Delta_s^2}\frac{g_{u,\ell}^2}{\Delta_{u,\ell}} \sim 1.28\Gamma'$, which could be achieved for $|\Omega_s|^2/\Delta_s^2 \sim 0.05$, $g_u \sim 2\pi\times 2$GHz, $\Delta_u\sim 2\pi\times 31$GHz and $\Gamma'\sim 2\pi\times 5$MHz. Furthermore the values of $g_u$ and $L_u$ are then consistent with a photonic crystal having $g_\lambda \sim 2\pi\times 10$GHz (such as for the ``alligator'' structure described in Ref.~\cite{Douglas2014a}) and band curvature $\alpha \sim 2$ (such as for the structure described in Ref.~\cite{Hung2013a}) for $z_a\sim\lambda$.


\begin{thebibliography}{48}%
\makeatletter
\providecommand \@ifxundefined [1]{%
 \@ifx{#1\undefined}
}%
\providecommand \@ifnum [1]{%
 \ifnum #1\expandafter \@firstoftwo
 \else \expandafter \@secondoftwo
 \fi
}%
\providecommand \@ifx [1]{%
 \ifx #1\expandafter \@firstoftwo
 \else \expandafter \@secondoftwo
 \fi
}%
\providecommand \natexlab [1]{#1}%
\providecommand \enquote  [1]{``#1''}%
\providecommand \bibnamefont  [1]{#1}%
\providecommand \bibfnamefont [1]{#1}%
\providecommand \citenamefont [1]{#1}%
\providecommand \href@noop [0]{\@secondoftwo}%
\providecommand \href [0]{\begingroup \@sanitize@url \@href}%
\providecommand \@href[1]{\@@startlink{#1}\@@href}%
\providecommand \@@href[1]{\endgroup#1\@@endlink}%
\providecommand \@sanitize@url [0]{\catcode `\\12\catcode `\$12\catcode
  `\&12\catcode `\#12\catcode `\^12\catcode `\_12\catcode `\%12\relax}%
\providecommand \@@startlink[1]{}%
\providecommand \@@endlink[0]{}%
\providecommand \url  [0]{\begingroup\@sanitize@url \@url }%
\providecommand \@url [1]{\endgroup\@href {#1}{\urlprefix }}%
\providecommand \urlprefix  [0]{URL }%
\providecommand \Eprint [0]{\href }%
\providecommand \doibase [0]{http://dx.doi.org/}%
\providecommand \selectlanguage [0]{\@gobble}%
\providecommand \bibinfo  [0]{\@secondoftwo}%
\providecommand \bibfield  [0]{\@secondoftwo}%
\providecommand \translation [1]{[#1]}%
\providecommand \BibitemOpen [0]{}%
\providecommand \bibitemStop [0]{}%
\providecommand \bibitemNoStop [0]{.\EOS\space}%
\providecommand \EOS [0]{\spacefactor3000\relax}%
\providecommand \BibitemShut  [1]{\csname bibitem#1\endcsname}%
\let\auto@bib@innerbib\@empty
\bibitem [{\citenamefont {Chang}\ \emph {et~al.}(2014)\citenamefont {Chang},
  \citenamefont {Vuletic},\ and\ \citenamefont {Lukin}}]{Chang2014a}%
  \BibitemOpen
  \bibfield  {author} {\bibinfo {author} {\bibfnamefont {D.~E.}\ \bibnamefont
  {Chang}}, \bibinfo {author} {\bibfnamefont {V.}~\bibnamefont {Vuletic}}, \
  and\ \bibinfo {author} {\bibfnamefont {M.~D.}\ \bibnamefont {Lukin}},\
  }\bibfield  {title} {\emph {\bibinfo {title} {Quantum nonlinear optics ---
  photon by photon},\ }}\href {http://dx.doi.org/10.1038/nphoton.2014.192}
  {\bibfield  {journal} {\bibinfo  {journal} {Nat. Photon.}\ }\textbf {\bibinfo
  {volume} {8}},\ \bibinfo {pages} {685} (\bibinfo {year} {2014})}\BibitemShut
  {NoStop}%
\bibitem [{\citenamefont {Turchette}\ \emph {et~al.}(1995)\citenamefont
  {Turchette}, \citenamefont {Hood}, \citenamefont {Lange}, \citenamefont
  {Mabuchi},\ and\ \citenamefont {Kimble}}]{Turchette1995a}%
  \BibitemOpen
  \bibfield  {author} {\bibinfo {author} {\bibfnamefont {Q.~A.}\ \bibnamefont
  {Turchette}}, \bibinfo {author} {\bibfnamefont {C.~J.}\ \bibnamefont {Hood}},
  \bibinfo {author} {\bibfnamefont {W.}~\bibnamefont {Lange}}, \bibinfo
  {author} {\bibfnamefont {H.}~\bibnamefont {Mabuchi}}, \ and\ \bibinfo
  {author} {\bibfnamefont {H.~J.}\ \bibnamefont {Kimble}},\ }\bibfield  {title}
  {\emph {\bibinfo {title} {Measurement of conditional phase shifts for quantum
  logic},\ }}\href {\doibase 10.1103/PhysRevLett.75.4710} {\bibfield  {journal}
  {\bibinfo  {journal} {Phys. Rev. Lett.}\ }\textbf {\bibinfo {volume} {75}},\
  \bibinfo {pages} {4710} (\bibinfo {year} {1995})}\BibitemShut {NoStop}%
\bibitem [{\citenamefont {Imamo\ifmmode~\bar{g}\else \={g}\fi{}lu}\ \emph
  {et~al.}(1997)\citenamefont {Imamo\ifmmode~\bar{g}\else \={g}\fi{}lu},
  \citenamefont {Schmidt}, \citenamefont {Woods},\ and\ \citenamefont
  {Deutsch}}]{Imamoglu1997c}%
  \BibitemOpen
  \bibfield  {author} {\bibinfo {author} {\bibfnamefont {A.}~\bibnamefont
  {Imamo\ifmmode~\bar{g}\else \={g}\fi{}lu}}, \bibinfo {author} {\bibfnamefont
  {H.}~\bibnamefont {Schmidt}}, \bibinfo {author} {\bibfnamefont
  {G.}~\bibnamefont {Woods}}, \ and\ \bibinfo {author} {\bibfnamefont
  {M.}~\bibnamefont {Deutsch}},\ }\bibfield  {title} {\emph {\bibinfo {title}
  {Strongly interacting photons in a nonlinear cavity},\ }}\href {\doibase
  10.1103/PhysRevLett.79.1467} {\bibfield  {journal} {\bibinfo  {journal}
  {Phys. Rev. Lett.}\ }\textbf {\bibinfo {volume} {79}},\ \bibinfo {pages}
  {1467} (\bibinfo {year} {1997})}\BibitemShut {NoStop}%
\bibitem [{\citenamefont {Duan}\ and\ \citenamefont
  {Kimble}(2004)}]{Duan2004a}%
  \BibitemOpen
  \bibfield  {author} {\bibinfo {author} {\bibfnamefont {L.-M.}\ \bibnamefont
  {Duan}}\ and\ \bibinfo {author} {\bibfnamefont {H.~J.}\ \bibnamefont
  {Kimble}},\ }\bibfield  {title} {\emph {\bibinfo {title} {Scalable photonic
  quantum computation through cavity-assisted interactions},\ }}\href {\doibase
  10.1103/PhysRevLett.92.127902} {\bibfield  {journal} {\bibinfo  {journal}
  {Phys. Rev. Lett.}\ }\textbf {\bibinfo {volume} {92}},\ \bibinfo {pages}
  {127902} (\bibinfo {year} {2004})}\BibitemShut {NoStop}%
\bibitem [{\citenamefont {Birnbaum}\ \emph {et~al.}(2005)\citenamefont
  {Birnbaum}, \citenamefont {Boca}, \citenamefont {Miller}, \citenamefont
  {Boozer}, \citenamefont {Northup},\ and\ \citenamefont
  {Kimble}}]{Birnbaum2005a}%
  \BibitemOpen
  \bibfield  {author} {\bibinfo {author} {\bibfnamefont {K.~M.}\ \bibnamefont
  {Birnbaum}}, \bibinfo {author} {\bibfnamefont {A.}~\bibnamefont {Boca}},
  \bibinfo {author} {\bibfnamefont {R.}~\bibnamefont {Miller}}, \bibinfo
  {author} {\bibfnamefont {A.~D.}\ \bibnamefont {Boozer}}, \bibinfo {author}
  {\bibfnamefont {T.~E.}\ \bibnamefont {Northup}}, \ and\ \bibinfo {author}
  {\bibfnamefont {H.~J.}\ \bibnamefont {Kimble}},\ }\bibfield  {title} {\emph
  {\bibinfo {title} {Photon blockade in an optical cavity with one trapped
  atom},\ }}\href {http://dx.doi.org/10.1038/nature03804} {\bibfield  {journal}
  {\bibinfo  {journal} {Nature}\ }\textbf {\bibinfo {volume} {436}},\ \bibinfo
  {pages} {87} (\bibinfo {year} {2005})}\BibitemShut {NoStop}%
\bibitem [{\citenamefont {Haroche}\ and\ \citenamefont
  {Raimond}(2006)}]{Haroche2006}%
  \BibitemOpen
  \bibfield  {author} {\bibinfo {author} {\bibfnamefont {S.}~\bibnamefont
  {Haroche}}\ and\ \bibinfo {author} {\bibfnamefont {J.~M.}\ \bibnamefont
  {Raimond}},\ }\href@noop {} {\emph {\bibinfo {title} {Exploring the
  quantum}}}\ (\bibinfo  {publisher} {Oxford Univ. Press},\ \bibinfo {year}
  {2006})\BibitemShut {NoStop}%
\bibitem [{\citenamefont {Kimble}(2008)}]{Kimble2008a}%
  \BibitemOpen
  \bibfield  {author} {\bibinfo {author} {\bibfnamefont {H.~J.}\ \bibnamefont
  {Kimble}},\ }\bibfield  {title} {\emph {\bibinfo {title} {The quantum
  internet},\ }}\href {http://dx.doi.org/10.1038/nature07127} {\bibfield
  {journal} {\bibinfo  {journal} {Nature}\ }\textbf {\bibinfo {volume} {453}},\
  \bibinfo {pages} {1023} (\bibinfo {year} {2008})}\BibitemShut {NoStop}%
\bibitem [{\citenamefont {{Reiserer}}\ and\ \citenamefont
  {{Rempe}}(2014)}]{Reiserer2015a}%
  \BibitemOpen
  \bibfield  {author} {\bibinfo {author} {\bibfnamefont {A.}~\bibnamefont
  {{Reiserer}}}\ and\ \bibinfo {author} {\bibfnamefont {G.}~\bibnamefont
  {{Rempe}}},\ }\bibfield  {title} {\emph {\bibinfo {title} {{Cavity-based
  quantum networks with single atoms and optical photons}},\ }}\href@noop {}
  {\bibfield  {journal} {\bibinfo  {journal} {ArXiv e-prints}\ } (\bibinfo
  {year} {2014})},\ \Eprint {http://arxiv.org/abs/1412.2889} {arXiv:1412.2889
  [quant-ph]} \BibitemShut {NoStop}%
\bibitem [{\citenamefont {Pritchard}\ \emph {et~al.}(2010)\citenamefont
  {Pritchard}, \citenamefont {Maxwell}, \citenamefont {Gauguet}, \citenamefont
  {Weatherill}, \citenamefont {Jones},\ and\ \citenamefont
  {Adams}}]{Pritchard2010a}%
  \BibitemOpen
  \bibfield  {author} {\bibinfo {author} {\bibfnamefont {J.~D.}\ \bibnamefont
  {Pritchard}}, \bibinfo {author} {\bibfnamefont {D.}~\bibnamefont {Maxwell}},
  \bibinfo {author} {\bibfnamefont {A.}~\bibnamefont {Gauguet}}, \bibinfo
  {author} {\bibfnamefont {K.~J.}\ \bibnamefont {Weatherill}}, \bibinfo
  {author} {\bibfnamefont {M.~P.~A.}\ \bibnamefont {Jones}}, \ and\ \bibinfo
  {author} {\bibfnamefont {C.~S.}\ \bibnamefont {Adams}},\ }\bibfield  {title}
  {\emph {\bibinfo {title} {Cooperative atom-light interaction in a blockaded
  {R}ydberg ensemble},\ }}\href {\doibase 10.1103/PhysRevLett.105.193603}
  {\bibfield  {journal} {\bibinfo  {journal} {Phys. Rev. Lett.}\ }\textbf
  {\bibinfo {volume} {105}},\ \bibinfo {pages} {193603} (\bibinfo {year}
  {2010})}\BibitemShut {NoStop}%
\bibitem [{\citenamefont {Dudin}\ and\ \citenamefont
  {Kuzmich}(2012)}]{Dudin2012a}%
  \BibitemOpen
  \bibfield  {author} {\bibinfo {author} {\bibfnamefont {Y.~O.}\ \bibnamefont
  {Dudin}}\ and\ \bibinfo {author} {\bibfnamefont {A.}~\bibnamefont
  {Kuzmich}},\ }\bibfield  {title} {\emph {\bibinfo {title} {Strongly
  interacting {R}ydberg excitations of a cold atomic gas},\ }}\href {\doibase
  10.1126/science.1217901} {\bibfield  {journal} {\bibinfo  {journal}
  {Science}\ }\textbf {\bibinfo {volume} {336}},\ \bibinfo {pages} {887}
  (\bibinfo {year} {2012})}\BibitemShut {NoStop}%
\bibitem [{\citenamefont {{Pritchard}}\ \emph {et~al.}(2013)\citenamefont
  {{Pritchard}}, \citenamefont {{Weatherill}},\ and\ \citenamefont
  {{Adams}}}]{Pritchard2013a}%
  \BibitemOpen
  \bibfield  {author} {\bibinfo {author} {\bibfnamefont {J.~D.}\ \bibnamefont
  {{Pritchard}}}, \bibinfo {author} {\bibfnamefont {K.~J.}\ \bibnamefont
  {{Weatherill}}}, \ and\ \bibinfo {author} {\bibfnamefont {C.~S.}\
  \bibnamefont {{Adams}}},\ }\bibinfo {title} {{Nonlinear Optics Using Cold
  Rydberg Atoms}},\ in\ \href {\doibase 10.1142/9789814440400_0008} {\emph
  {\bibinfo {booktitle} {Annual Review of Cold Atoms and Molecules, Volume
  1.}}},\ \bibinfo {editor} {edited by\ \bibinfo {editor} {\bibfnamefont
  {K.}~\bibnamefont {{Madison}}}, \bibinfo {editor} {\bibfnamefont
  {Y.}~\bibnamefont {Wang}}, \ and\ \bibinfo {editor} {\bibfnamefont {A.~M.}\
  \bibnamefont {Rey}}}\ (\bibinfo  {publisher} {World Scientific Publishing
  Co},\ \bibinfo {year} {2013})\ pp.\ \bibinfo {pages} {301--350}\BibitemShut
  {NoStop}%
\bibitem [{\citenamefont {Peyronel}\ \emph {et~al.}(2012)\citenamefont
  {Peyronel}, \citenamefont {Firstenberg}, \citenamefont {Liang}, \citenamefont
  {Hofferberth}, \citenamefont {Gorshkov}, \citenamefont {Pohl}, \citenamefont
  {Lukin},\ and\ \citenamefont {Vuletic}}]{Peyronel2012a}%
  \BibitemOpen
  \bibfield  {author} {\bibinfo {author} {\bibfnamefont {T.}~\bibnamefont
  {Peyronel}}, \bibinfo {author} {\bibfnamefont {O.}~\bibnamefont
  {Firstenberg}}, \bibinfo {author} {\bibfnamefont {Q.-Y.}\ \bibnamefont
  {Liang}}, \bibinfo {author} {\bibfnamefont {S.}~\bibnamefont {Hofferberth}},
  \bibinfo {author} {\bibfnamefont {A.~V.}\ \bibnamefont {Gorshkov}}, \bibinfo
  {author} {\bibfnamefont {T.}~\bibnamefont {Pohl}}, \bibinfo {author}
  {\bibfnamefont {M.~D.}\ \bibnamefont {Lukin}}, \ and\ \bibinfo {author}
  {\bibfnamefont {V.}~\bibnamefont {Vuletic}},\ }\bibfield  {title} {\emph
  {\bibinfo {title} {Quantum nonlinear optics with single photons enabled by
  strongly interacting atoms},\ }}\href {http://dx.doi.org/10.1038/nature11361}
  {\bibfield  {journal} {\bibinfo  {journal} {Nature}\ }\textbf {\bibinfo
  {volume} {488}},\ \bibinfo {pages} {57} (\bibinfo {year} {2012})}\BibitemShut
  {NoStop}%
\bibitem [{\citenamefont {Baur}\ \emph {et~al.}(2014)\citenamefont {Baur},
  \citenamefont {Tiarks}, \citenamefont {Rempe},\ and\ \citenamefont
  {D\"urr}}]{Baur2014a}%
  \BibitemOpen
  \bibfield  {author} {\bibinfo {author} {\bibfnamefont {S.}~\bibnamefont
  {Baur}}, \bibinfo {author} {\bibfnamefont {D.}~\bibnamefont {Tiarks}},
  \bibinfo {author} {\bibfnamefont {G.}~\bibnamefont {Rempe}}, \ and\ \bibinfo
  {author} {\bibfnamefont {S.}~\bibnamefont {D\"urr}},\ }\bibfield  {title}
  {\emph {\bibinfo {title} {Single-photon switch based on rydberg blockade},\
  }}\href {\doibase 10.1103/PhysRevLett.112.073901} {\bibfield  {journal}
  {\bibinfo  {journal} {Phys. Rev. Lett.}\ }\textbf {\bibinfo {volume} {112}},\
  \bibinfo {pages} {073901} (\bibinfo {year} {2014})}\BibitemShut {NoStop}%
\bibitem [{\citenamefont {Gorniaczyk}\ \emph {et~al.}(2014)\citenamefont
  {Gorniaczyk}, \citenamefont {Tresp}, \citenamefont {Schmidt}, \citenamefont
  {Fedder},\ and\ \citenamefont {Hofferberth}}]{Gorniaczyk2014a}%
  \BibitemOpen
  \bibfield  {author} {\bibinfo {author} {\bibfnamefont {H.}~\bibnamefont
  {Gorniaczyk}}, \bibinfo {author} {\bibfnamefont {C.}~\bibnamefont {Tresp}},
  \bibinfo {author} {\bibfnamefont {J.}~\bibnamefont {Schmidt}}, \bibinfo
  {author} {\bibfnamefont {H.}~\bibnamefont {Fedder}}, \ and\ \bibinfo {author}
  {\bibfnamefont {S.}~\bibnamefont {Hofferberth}},\ }\bibfield  {title} {\emph
  {\bibinfo {title} {Single-photon transistor mediated by interstate {R}ydberg
  interactions},\ }}\href {\doibase 10.1103/PhysRevLett.113.053601} {\bibfield
  {journal} {\bibinfo  {journal} {Phys. Rev. Lett.}\ }\textbf {\bibinfo
  {volume} {113}},\ \bibinfo {pages} {053601} (\bibinfo {year}
  {2014})}\BibitemShut {NoStop}%
\bibitem [{\citenamefont {Tiarks}\ \emph {et~al.}(2014)\citenamefont {Tiarks},
  \citenamefont {Baur}, \citenamefont {Schneider}, \citenamefont {D\"urr},\
  and\ \citenamefont {Rempe}}]{Tiarks2014a}%
  \BibitemOpen
  \bibfield  {author} {\bibinfo {author} {\bibfnamefont {D.}~\bibnamefont
  {Tiarks}}, \bibinfo {author} {\bibfnamefont {S.}~\bibnamefont {Baur}},
  \bibinfo {author} {\bibfnamefont {K.}~\bibnamefont {Schneider}}, \bibinfo
  {author} {\bibfnamefont {S.}~\bibnamefont {D\"urr}}, \ and\ \bibinfo {author}
  {\bibfnamefont {G.}~\bibnamefont {Rempe}},\ }\bibfield  {title} {\emph
  {\bibinfo {title} {Single-photon transistor using a {F}\"orster resonance},\
  }}\href {\doibase 10.1103/PhysRevLett.113.053602} {\bibfield  {journal}
  {\bibinfo  {journal} {Phys. Rev. Lett.}\ }\textbf {\bibinfo {volume} {113}},\
  \bibinfo {pages} {053602} (\bibinfo {year} {2014})}\BibitemShut {NoStop}%
\bibitem [{\citenamefont {Gorshkov}\ \emph {et~al.}(2011)\citenamefont
  {Gorshkov}, \citenamefont {Otterbach}, \citenamefont {Fleischhauer},
  \citenamefont {Pohl},\ and\ \citenamefont {Lukin}}]{Gorshkov2011a}%
  \BibitemOpen
  \bibfield  {author} {\bibinfo {author} {\bibfnamefont {A.~V.}\ \bibnamefont
  {Gorshkov}}, \bibinfo {author} {\bibfnamefont {J.}~\bibnamefont {Otterbach}},
  \bibinfo {author} {\bibfnamefont {M.}~\bibnamefont {Fleischhauer}}, \bibinfo
  {author} {\bibfnamefont {T.}~\bibnamefont {Pohl}}, \ and\ \bibinfo {author}
  {\bibfnamefont {M.~D.}\ \bibnamefont {Lukin}},\ }\bibfield  {title} {\emph
  {\bibinfo {title} {Photon-photon interactions via {R}ydberg blockade},\
  }}\href {\doibase 10.1103/PhysRevLett.107.133602} {\bibfield  {journal}
  {\bibinfo  {journal} {Phys. Rev. Lett.}\ }\textbf {\bibinfo {volume} {107}},\
  \bibinfo {pages} {133602} (\bibinfo {year} {2011})}\BibitemShut {NoStop}%
\bibitem [{\citenamefont {Firstenberg}\ \emph {et~al.}(2013)\citenamefont
  {Firstenberg}, \citenamefont {Peyronel}, \citenamefont {Liang}, \citenamefont
  {Gorshkov}, \citenamefont {Lukin},\ and\ \citenamefont
  {Vuletic}}]{Firstenberg2013a}%
  \BibitemOpen
  \bibfield  {author} {\bibinfo {author} {\bibfnamefont {O.}~\bibnamefont
  {Firstenberg}}, \bibinfo {author} {\bibfnamefont {T.}~\bibnamefont
  {Peyronel}}, \bibinfo {author} {\bibfnamefont {Q.-Y.}\ \bibnamefont {Liang}},
  \bibinfo {author} {\bibfnamefont {A.~V.}\ \bibnamefont {Gorshkov}}, \bibinfo
  {author} {\bibfnamefont {M.~D.}\ \bibnamefont {Lukin}}, \ and\ \bibinfo
  {author} {\bibfnamefont {V.}~\bibnamefont {Vuletic}},\ }\bibfield  {title}
  {\emph {\bibinfo {title} {Attractive photons in a quantum nonlinear medium},\
  }}\href {http://dx.doi.org/10.1038/nature12512} {\bibfield  {journal}
  {\bibinfo  {journal} {Nature}\ }\textbf {\bibinfo {volume} {502}},\ \bibinfo
  {pages} {71} (\bibinfo {year} {2013})}\BibitemShut {NoStop}%
\bibitem [{\citenamefont {Otterbach}\ \emph {et~al.}(2013)\citenamefont
  {Otterbach}, \citenamefont {Moos}, \citenamefont {Muth},\ and\ \citenamefont
  {Fleischhauer}}]{Otterbach2013a}%
  \BibitemOpen
  \bibfield  {author} {\bibinfo {author} {\bibfnamefont {J.}~\bibnamefont
  {Otterbach}}, \bibinfo {author} {\bibfnamefont {M.}~\bibnamefont {Moos}},
  \bibinfo {author} {\bibfnamefont {D.}~\bibnamefont {Muth}}, \ and\ \bibinfo
  {author} {\bibfnamefont {M.}~\bibnamefont {Fleischhauer}},\ }\bibfield
  {title} {\emph {\bibinfo {title} {Wigner crystallization of single photons in
  cold {R}ydberg ensembles},\ }}\href {\doibase 10.1103/PhysRevLett.111.113001}
  {\bibfield  {journal} {\bibinfo  {journal} {Phys. Rev. Lett.}\ }\textbf
  {\bibinfo {volume} {111}},\ \bibinfo {pages} {113001} (\bibinfo {year}
  {2013})}\BibitemShut {NoStop}%
\bibitem [{\citenamefont {Bienias}\ \emph {et~al.}(2014)\citenamefont
  {Bienias}, \citenamefont {Choi}, \citenamefont {Firstenberg}, \citenamefont
  {Maghrebi}, \citenamefont {Gullans}, \citenamefont {Lukin}, \citenamefont
  {Gorshkov},\ and\ \citenamefont {B\"uchler}}]{Bienias2014a}%
  \BibitemOpen
  \bibfield  {author} {\bibinfo {author} {\bibfnamefont {P.}~\bibnamefont
  {Bienias}}, \bibinfo {author} {\bibfnamefont {S.}~\bibnamefont {Choi}},
  \bibinfo {author} {\bibfnamefont {O.}~\bibnamefont {Firstenberg}}, \bibinfo
  {author} {\bibfnamefont {M.~F.}\ \bibnamefont {Maghrebi}}, \bibinfo {author}
  {\bibfnamefont {M.}~\bibnamefont {Gullans}}, \bibinfo {author} {\bibfnamefont
  {M.~D.}\ \bibnamefont {Lukin}}, \bibinfo {author} {\bibfnamefont {A.~V.}\
  \bibnamefont {Gorshkov}}, \ and\ \bibinfo {author} {\bibfnamefont {H.~P.}\
  \bibnamefont {B\"uchler}},\ }\bibfield  {title} {\emph {\bibinfo {title}
  {Scattering resonances and bound states for strongly interacting {R}ydberg
  polaritons},\ }}\href {\doibase 10.1103/PhysRevA.90.053804} {\bibfield
  {journal} {\bibinfo  {journal} {Phys. Rev. A}\ }\textbf {\bibinfo {volume}
  {90}},\ \bibinfo {pages} {053804} (\bibinfo {year} {2014})}\BibitemShut
  {NoStop}%
\bibitem [{\citenamefont {Maghrebi}\ \emph {et~al.}(2015)\citenamefont
  {Maghrebi}, \citenamefont {Gullans}, \citenamefont {Bienias}, \citenamefont
  {Choi}, \citenamefont {Martin}, \citenamefont {Firstenberg}, \citenamefont
  {Lukin}, \citenamefont {B\"uchler},\ and\ \citenamefont
  {Gorshkov}}]{Maghrebi2015a}%
  \BibitemOpen
  \bibfield  {author} {\bibinfo {author} {\bibfnamefont {M.~F.}\ \bibnamefont
  {Maghrebi}}, \bibinfo {author} {\bibfnamefont {M.~J.}\ \bibnamefont
  {Gullans}}, \bibinfo {author} {\bibfnamefont {P.}~\bibnamefont {Bienias}},
  \bibinfo {author} {\bibfnamefont {S.}~\bibnamefont {Choi}}, \bibinfo {author}
  {\bibfnamefont {I.}~\bibnamefont {Martin}}, \bibinfo {author} {\bibfnamefont
  {O.}~\bibnamefont {Firstenberg}}, \bibinfo {author} {\bibfnamefont {M.~D.}\
  \bibnamefont {Lukin}}, \bibinfo {author} {\bibfnamefont {H.~P.}\ \bibnamefont
  {B\"uchler}}, \ and\ \bibinfo {author} {\bibfnamefont {A.~V.}\ \bibnamefont
  {Gorshkov}},\ }\bibfield  {title} {\emph {\bibinfo {title} {Coulomb bound
  states of strongly interacting photons},\ }}\href {\doibase
  10.1103/PhysRevLett.115.123601} {\bibfield  {journal} {\bibinfo  {journal}
  {Phys. Rev. Lett.}\ }\textbf {\bibinfo {volume} {115}},\ \bibinfo {pages}
  {123601} (\bibinfo {year} {2015})}\BibitemShut {NoStop}%
\bibitem [{\citenamefont {Saffman}\ \emph {et~al.}(2010)\citenamefont
  {Saffman}, \citenamefont {Walker},\ and\ \citenamefont
  {M\o{}lmer}}]{Saffman2010a}%
  \BibitemOpen
  \bibfield  {author} {\bibinfo {author} {\bibfnamefont {M.}~\bibnamefont
  {Saffman}}, \bibinfo {author} {\bibfnamefont {T.~G.}\ \bibnamefont {Walker}},
  \ and\ \bibinfo {author} {\bibfnamefont {K.}~\bibnamefont {M\o{}lmer}},\
  }\bibfield  {title} {\emph {\bibinfo {title} {Quantum information with
  {R}ydberg atoms},\ }}\href {\doibase 10.1103/RevModPhys.82.2313} {\bibfield
  {journal} {\bibinfo  {journal} {Rev. Mod. Phys.}\ }\textbf {\bibinfo {volume}
  {82}},\ \bibinfo {pages} {2313} (\bibinfo {year} {2010})}\BibitemShut
  {NoStop}%
\bibitem [{\citenamefont {Thompson}\ \emph {et~al.}(2013)\citenamefont
  {Thompson}, \citenamefont {Tiecke}, \citenamefont {de~Leon}, \citenamefont
  {Feist}, \citenamefont {Akimov}, \citenamefont {Gullans}, \citenamefont
  {Zibrov}, \citenamefont {Vuleti\'c},\ and\ \citenamefont
  {Lukin}}]{Thompson2013a}%
  \BibitemOpen
  \bibfield  {author} {\bibinfo {author} {\bibfnamefont {J.~D.}\ \bibnamefont
  {Thompson}}, \bibinfo {author} {\bibfnamefont {T.~G.}\ \bibnamefont
  {Tiecke}}, \bibinfo {author} {\bibfnamefont {N.~P.}\ \bibnamefont {de~Leon}},
  \bibinfo {author} {\bibfnamefont {J.}~\bibnamefont {Feist}}, \bibinfo
  {author} {\bibfnamefont {A.~V.}\ \bibnamefont {Akimov}}, \bibinfo {author}
  {\bibfnamefont {M.}~\bibnamefont {Gullans}}, \bibinfo {author} {\bibfnamefont
  {A.~S.}\ \bibnamefont {Zibrov}}, \bibinfo {author} {\bibfnamefont
  {V.}~\bibnamefont {Vuleti\'c}}, \ and\ \bibinfo {author} {\bibfnamefont
  {M.~D.}\ \bibnamefont {Lukin}},\ }\bibfield  {title} {\emph {\bibinfo {title}
  {Coupling a single trapped atom to a nanoscale optical cavity},\ }}\href
  {\doibase 10.1126/science.1237125} {\bibfield  {journal} {\bibinfo  {journal}
  {Science}\ }\textbf {\bibinfo {volume} {340}},\ \bibinfo {pages} {1202}
  (\bibinfo {year} {2013})}\BibitemShut {NoStop}%
\bibitem [{\citenamefont {Goban}\ \emph {et~al.}(2014)\citenamefont {Goban},
  \citenamefont {Hung}, \citenamefont {Yu}, \citenamefont {Hood}, \citenamefont
  {Muniz}, \citenamefont {Lee}, \citenamefont {Martin}, \citenamefont
  {McClung}, \citenamefont {Choi}, \citenamefont {Chang}, \citenamefont
  {Painter},\ and\ \citenamefont {Kimble}}]{Goban2013a}%
  \BibitemOpen
  \bibfield  {author} {\bibinfo {author} {\bibfnamefont {A.}~\bibnamefont
  {Goban}}, \bibinfo {author} {\bibfnamefont {C.-L.}\ \bibnamefont {Hung}},
  \bibinfo {author} {\bibfnamefont {S.-P.}\ \bibnamefont {Yu}}, \bibinfo
  {author} {\bibfnamefont {J.}~\bibnamefont {Hood}}, \bibinfo {author}
  {\bibfnamefont {J.}~\bibnamefont {Muniz}}, \bibinfo {author} {\bibfnamefont
  {J.}~\bibnamefont {Lee}}, \bibinfo {author} {\bibfnamefont {M.}~\bibnamefont
  {Martin}}, \bibinfo {author} {\bibfnamefont {A.}~\bibnamefont {McClung}},
  \bibinfo {author} {\bibfnamefont {K.}~\bibnamefont {Choi}}, \bibinfo {author}
  {\bibfnamefont {D.}~\bibnamefont {Chang}}, \bibinfo {author} {\bibfnamefont
  {O.}~\bibnamefont {Painter}}, \ and\ \bibinfo {author} {\bibfnamefont
  {H.}~\bibnamefont {Kimble}},\ }\bibfield  {title} {\emph {\bibinfo {title}
  {Atom-light interactions in photonic crystals},\ }}\href
  {http://dx.doi.org/10.1038/ncomms4808} {\bibfield  {journal} {\bibinfo
  {journal} {Nat. Commun.}\ }\textbf {\bibinfo {volume} {5}},\ \bibinfo {pages}
  {3808} (\bibinfo {year} {2014})}\BibitemShut {NoStop}%
\bibitem [{\citenamefont {Yu}\ \emph {et~al.}(2014)\citenamefont {Yu},
  \citenamefont {Hood}, \citenamefont {Muniz}, \citenamefont {Martin},
  \citenamefont {Norte}, \citenamefont {Hung}, \citenamefont {Meenehan},
  \citenamefont {Cohen}, \citenamefont {Painter},\ and\ \citenamefont
  {Kimble}}]{Yu2013a}%
  \BibitemOpen
  \bibfield  {author} {\bibinfo {author} {\bibfnamefont {S.-P.}\ \bibnamefont
  {Yu}}, \bibinfo {author} {\bibfnamefont {J.~D.}\ \bibnamefont {Hood}},
  \bibinfo {author} {\bibfnamefont {J.~A.}\ \bibnamefont {Muniz}}, \bibinfo
  {author} {\bibfnamefont {M.~J.}\ \bibnamefont {Martin}}, \bibinfo {author}
  {\bibfnamefont {R.}~\bibnamefont {Norte}}, \bibinfo {author} {\bibfnamefont
  {C.-L.}\ \bibnamefont {Hung}}, \bibinfo {author} {\bibfnamefont {S.~M.}\
  \bibnamefont {Meenehan}}, \bibinfo {author} {\bibfnamefont {J.~D.}\
  \bibnamefont {Cohen}}, \bibinfo {author} {\bibfnamefont {O.}~\bibnamefont
  {Painter}}, \ and\ \bibinfo {author} {\bibfnamefont {H.~J.}\ \bibnamefont
  {Kimble}},\ }\bibfield  {title} {\emph {\bibinfo {title} {Nanowire photonic
  crystal waveguides for single-atom trapping and strong light-matter
  interactions},\ }}\href {\doibase http://dx.doi.org/10.1063/1.4868975}
  {\bibfield  {journal} {\bibinfo  {journal} {Applied Physics Letters}\
  }\textbf {\bibinfo {volume} {104}},\ \bibinfo {eid} {111103} (\bibinfo {year}
  {2014})}\BibitemShut {NoStop}%
\bibitem [{\citenamefont {Tiecke}\ \emph {et~al.}(2014)\citenamefont {Tiecke},
  \citenamefont {Thompson}, \citenamefont {de~Leon}, \citenamefont {Liu},
  \citenamefont {Vuletic},\ and\ \citenamefont {Lukin}}]{Tiecke2014a}%
  \BibitemOpen
  \bibfield  {author} {\bibinfo {author} {\bibfnamefont {T.~G.}\ \bibnamefont
  {Tiecke}}, \bibinfo {author} {\bibfnamefont {J.~D.}\ \bibnamefont
  {Thompson}}, \bibinfo {author} {\bibfnamefont {N.~P.}\ \bibnamefont
  {de~Leon}}, \bibinfo {author} {\bibfnamefont {L.~R.}\ \bibnamefont {Liu}},
  \bibinfo {author} {\bibfnamefont {V.}~\bibnamefont {Vuletic}}, \ and\
  \bibinfo {author} {\bibfnamefont {M.~D.}\ \bibnamefont {Lukin}},\ }\bibfield
  {title} {\emph {\bibinfo {title} {Nanophotonic quantum phase switch with a
  single atom},\ }}\href {http://dx.doi.org/10.1038/nature13188} {\bibfield
  {journal} {\bibinfo  {journal} {Nature}\ }\textbf {\bibinfo {volume} {508}},\
  \bibinfo {pages} {241} (\bibinfo {year} {2014})}\BibitemShut {NoStop}%
\bibitem [{\citenamefont {Goban}\ \emph {et~al.}(2015)\citenamefont {Goban},
  \citenamefont {Hung}, \citenamefont {Hood}, \citenamefont {Yu}, \citenamefont
  {Muniz}, \citenamefont {Painter},\ and\ \citenamefont {Kimble}}]{Goban2015a}%
  \BibitemOpen
  \bibfield  {author} {\bibinfo {author} {\bibfnamefont {A.}~\bibnamefont
  {Goban}}, \bibinfo {author} {\bibfnamefont {C.-L.}\ \bibnamefont {Hung}},
  \bibinfo {author} {\bibfnamefont {J.~D.}\ \bibnamefont {Hood}}, \bibinfo
  {author} {\bibfnamefont {S.-P.}\ \bibnamefont {Yu}}, \bibinfo {author}
  {\bibfnamefont {J.~A.}\ \bibnamefont {Muniz}}, \bibinfo {author}
  {\bibfnamefont {O.}~\bibnamefont {Painter}}, \ and\ \bibinfo {author}
  {\bibfnamefont {H.~J.}\ \bibnamefont {Kimble}},\ }\bibfield  {title} {\emph
  {\bibinfo {title} {Superradiance for atoms trapped along a photonic crystal
  waveguide},\ }}\href {\doibase 10.1103/PhysRevLett.115.063601} {\bibfield
  {journal} {\bibinfo  {journal} {Phys. Rev. Lett.}\ }\textbf {\bibinfo
  {volume} {115}},\ \bibinfo {pages} {063601} (\bibinfo {year}
  {2015})}\BibitemShut {NoStop}%
\bibitem [{\citenamefont {Kurizki}(1990)}]{Kurizki1990a}%
  \BibitemOpen
  \bibfield  {author} {\bibinfo {author} {\bibfnamefont {G.}~\bibnamefont
  {Kurizki}},\ }\bibfield  {title} {\emph {\bibinfo {title} {Two-atom resonant
  radiative coupling in photonic band structures},\ }}\href {\doibase
  10.1103/PhysRevA.42.2915} {\bibfield  {journal} {\bibinfo  {journal} {Phys.
  Rev. A}\ }\textbf {\bibinfo {volume} {42}},\ \bibinfo {pages} {2915}
  (\bibinfo {year} {1990})}\BibitemShut {NoStop}%
\bibitem [{\citenamefont {John}\ and\ \citenamefont {Wang}(1991)}]{John1991a}%
  \BibitemOpen
  \bibfield  {author} {\bibinfo {author} {\bibfnamefont {S.}~\bibnamefont
  {John}}\ and\ \bibinfo {author} {\bibfnamefont {J.}~\bibnamefont {Wang}},\
  }\bibfield  {title} {\emph {\bibinfo {title} {Quantum optics of localized
  light in a photonic band gap},\ }}\href {\doibase 10.1103/PhysRevB.43.12772}
  {\bibfield  {journal} {\bibinfo  {journal} {Phys. Rev. B}\ }\textbf {\bibinfo
  {volume} {43}},\ \bibinfo {pages} {12772} (\bibinfo {year}
  {1991})}\BibitemShut {NoStop}%
\bibitem [{\citenamefont {Shahmoon}\ and\ \citenamefont
  {Kurizki}(2013)}]{Shahmoon2013a}%
  \BibitemOpen
  \bibfield  {author} {\bibinfo {author} {\bibfnamefont {E.}~\bibnamefont
  {Shahmoon}}\ and\ \bibinfo {author} {\bibfnamefont {G.}~\bibnamefont
  {Kurizki}},\ }\bibfield  {title} {\emph {\bibinfo {title} {Nonradiative
  interaction and entanglement between distant atoms},\ }}\href {\doibase
  10.1103/PhysRevA.87.033831} {\bibfield  {journal} {\bibinfo  {journal} {Phys.
  Rev. A}\ }\textbf {\bibinfo {volume} {87}},\ \bibinfo {pages} {033831}
  (\bibinfo {year} {2013})}\BibitemShut {NoStop}%
\bibitem [{\citenamefont {{Douglas}}\ \emph {et~al.}(2015)\citenamefont
  {{Douglas}}, \citenamefont {{Habibian}}, \citenamefont {{Hung}},
  \citenamefont {{Gorshkov}}, \citenamefont {{Kimble}},\ and\ \citenamefont
  {{Chang}}}]{Douglas2014a}%
  \BibitemOpen
  \bibfield  {author} {\bibinfo {author} {\bibfnamefont {J.~S.}\ \bibnamefont
  {{Douglas}}}, \bibinfo {author} {\bibfnamefont {H.}~\bibnamefont
  {{Habibian}}}, \bibinfo {author} {\bibfnamefont {C.-L.}\ \bibnamefont
  {{Hung}}}, \bibinfo {author} {\bibfnamefont {A.~V.}\ \bibnamefont
  {{Gorshkov}}}, \bibinfo {author} {\bibfnamefont {H.~J.}\ \bibnamefont
  {{Kimble}}}, \ and\ \bibinfo {author} {\bibfnamefont {D.~E.}\ \bibnamefont
  {{Chang}}},\ }\bibfield  {title} {\emph {\bibinfo {title} {Quantum many-body
  models with cold atoms coupled to photonic crystals},\ }}\href
  {http://dx.doi.org/10.1038/nphoton.2015.57} {\bibfield  {journal} {\bibinfo
  {journal} {Nat. Photon.}\ }\textbf {\bibinfo {volume} {9}},\ \bibinfo {pages}
  {326} (\bibinfo {year} {2015})}\BibitemShut {NoStop}%
\bibitem [{\citenamefont {{Gonz{\'a}lez-Tudela}}\ \emph
  {et~al.}(2015)\citenamefont {{Gonz{\'a}lez-Tudela}}, \citenamefont {{Hung}},
  \citenamefont {{Chang}}, \citenamefont {{Cirac}},\ and\ \citenamefont
  {{Kimble}}}]{Tudela2014a}%
  \BibitemOpen
  \bibfield  {author} {\bibinfo {author} {\bibfnamefont {A.}~\bibnamefont
  {{Gonz{\'a}lez-Tudela}}}, \bibinfo {author} {\bibfnamefont {C.-L.}\
  \bibnamefont {{Hung}}}, \bibinfo {author} {\bibfnamefont {D.~E.}\
  \bibnamefont {{Chang}}}, \bibinfo {author} {\bibfnamefont {J.~I.}\
  \bibnamefont {{Cirac}}}, \ and\ \bibinfo {author} {\bibfnamefont {H.~J.}\
  \bibnamefont {{Kimble}}},\ }\bibfield  {title} {\emph {\bibinfo {title}
  {{Subwavelength vacuum lattices and atom-atom interactions in two-dimensional
  photonic crystals}},\ }}\href {http://dx.doi.org/10.1038/nphoton.2015.54}
  {\bibfield  {journal} {\bibinfo  {journal} {Nat Photon}\ }\textbf {\bibinfo
  {volume} {9}},\ \bibinfo {pages} {320} (\bibinfo {year} {2015})}\BibitemShut
  {NoStop}%
\bibitem [{\citenamefont {{Shahmoon}}\ \emph {et~al.}(2014)\citenamefont
  {{Shahmoon}}, \citenamefont {{Grisins}}, \citenamefont {{Stimming}},
  \citenamefont {{Mazets}},\ and\ \citenamefont {{Kurizki}}}]{Shahmoon2015a}%
  \BibitemOpen
  \bibfield  {author} {\bibinfo {author} {\bibfnamefont {E.}~\bibnamefont
  {{Shahmoon}}}, \bibinfo {author} {\bibfnamefont {P.}~\bibnamefont
  {{Grisins}}}, \bibinfo {author} {\bibfnamefont {H.~P.}\ \bibnamefont
  {{Stimming}}}, \bibinfo {author} {\bibfnamefont {I.}~\bibnamefont
  {{Mazets}}}, \ and\ \bibinfo {author} {\bibfnamefont {G.}~\bibnamefont
  {{Kurizki}}},\ }\bibfield  {title} {\emph {\bibinfo {title} {{Extremely
  nonlocal optical nonlinearities in atoms trapped near a waveguide}},\
  }}\href@noop {} {\bibfield  {journal} {\bibinfo  {journal} {ArXiv e-prints}\
  } (\bibinfo {year} {2014})},\ \Eprint {http://arxiv.org/abs/1412.8331}
  {arXiv:1412.8331 [quant-ph]} \BibitemShut {NoStop}%
\bibitem [{\citenamefont {Fleischhauer}\ \emph {et~al.}(2005)\citenamefont
  {Fleischhauer}, \citenamefont {Imamoglu},\ and\ \citenamefont
  {Marangos}}]{Fleischhauer2005a}%
  \BibitemOpen
  \bibfield  {author} {\bibinfo {author} {\bibfnamefont {M.}~\bibnamefont
  {Fleischhauer}}, \bibinfo {author} {\bibfnamefont {A.}~\bibnamefont
  {Imamoglu}}, \ and\ \bibinfo {author} {\bibfnamefont {J.~P.}\ \bibnamefont
  {Marangos}},\ }\bibfield  {title} {\emph {\bibinfo {title}
  {Electromagnetically induced transparency: Optics in coherent media},\
  }}\href {\doibase 10.1103/RevModPhys.77.633} {\bibfield  {journal} {\bibinfo
  {journal} {Rev. Mod. Phys.}\ }\textbf {\bibinfo {volume} {77}},\ \bibinfo
  {pages} {633} (\bibinfo {year} {2005})}\BibitemShut {NoStop}%
\bibitem [{\citenamefont {Joannopoulos}\ \emph {et~al.}(2008)\citenamefont
  {Joannopoulos}, \citenamefont {Johnson}, \citenamefont {Winn},\ and\
  \citenamefont {Meade}}]{Joannopoulos2008}%
  \BibitemOpen
  \bibfield  {author} {\bibinfo {author} {\bibfnamefont {J.~D.}\ \bibnamefont
  {Joannopoulos}}, \bibinfo {author} {\bibfnamefont {S.~G.}\ \bibnamefont
  {Johnson}}, \bibinfo {author} {\bibfnamefont {J.~N.}\ \bibnamefont {Winn}}, \
  and\ \bibinfo {author} {\bibfnamefont {R.~D.}\ \bibnamefont {Meade}},\
  }\href@noop {} {\emph {\bibinfo {title} {Photonic Crystals: Molding the Flow
  of Light (Second Edition)}}},\ \bibinfo {edition} {2nd}\ ed.\ (\bibinfo
  {publisher} {Princeton University Press},\ \bibinfo {year}
  {2008})\BibitemShut {NoStop}%
\bibitem [{\citenamefont {John}\ and\ \citenamefont {Wang}(1990)}]{John1990a}%
  \BibitemOpen
  \bibfield  {author} {\bibinfo {author} {\bibfnamefont {S.}~\bibnamefont
  {John}}\ and\ \bibinfo {author} {\bibfnamefont {J.}~\bibnamefont {Wang}},\
  }\bibfield  {title} {\emph {\bibinfo {title} {Quantum electrodynamics near a
  photonic band gap: Photon bound states and dressed atoms},\ }}\href {\doibase
  10.1103/PhysRevLett.64.2418} {\bibfield  {journal} {\bibinfo  {journal}
  {Phys. Rev. Lett.}\ }\textbf {\bibinfo {volume} {64}},\ \bibinfo {pages}
  {2418} (\bibinfo {year} {1990})}\BibitemShut {NoStop}%
\bibitem [{\citenamefont {Agarwal}\ \emph {et~al.}(1995)\citenamefont
  {Agarwal}, \citenamefont {Gupta},\ and\ \citenamefont {Puri}}]{Agarwal1995}%
  \BibitemOpen
  \bibfield  {author} {\bibinfo {author} {\bibfnamefont {G.}~\bibnamefont
  {Agarwal}}, \bibinfo {author} {\bibfnamefont {S.}~\bibnamefont {Gupta}}, \
  and\ \bibinfo {author} {\bibfnamefont {R.}~\bibnamefont {Puri}},\ }\href
  {http://books.google.es/books?id=AJaMkQEACAAJ} {\emph {\bibinfo {title}
  {Fundamentals of Cavity Quantum Electrodynamics}}}\ (\bibinfo  {publisher}
  {World Scientific Publishing Company Incorporated},\ \bibinfo {year}
  {1995})\BibitemShut {NoStop}%
\bibitem [{\citenamefont {Bay}\ \emph {et~al.}(1997)\citenamefont {Bay},
  \citenamefont {Lambropoulos},\ and\ \citenamefont {M\o{}lmer}}]{Bay1997a}%
  \BibitemOpen
  \bibfield  {author} {\bibinfo {author} {\bibfnamefont {S.}~\bibnamefont
  {Bay}}, \bibinfo {author} {\bibfnamefont {P.}~\bibnamefont {Lambropoulos}}, \
  and\ \bibinfo {author} {\bibfnamefont {K.}~\bibnamefont {M\o{}lmer}},\
  }\bibfield  {title} {\emph {\bibinfo {title} {Atom-atom interaction in
  strongly modified reservoirs},\ }}\href {\doibase 10.1103/PhysRevA.55.1485}
  {\bibfield  {journal} {\bibinfo  {journal} {Phys. Rev. A}\ }\textbf {\bibinfo
  {volume} {55}},\ \bibinfo {pages} {1485} (\bibinfo {year}
  {1997})}\BibitemShut {NoStop}%
\bibitem [{\citenamefont {Lambropoulos}\ \emph {et~al.}(2000)\citenamefont
  {Lambropoulos}, \citenamefont {Nikolopoulos}, \citenamefont {Nielsen},\ and\
  \citenamefont {Bay}}]{Lambropoulos2000a}%
  \BibitemOpen
  \bibfield  {author} {\bibinfo {author} {\bibfnamefont {P.}~\bibnamefont
  {Lambropoulos}}, \bibinfo {author} {\bibfnamefont {G.~M.}\ \bibnamefont
  {Nikolopoulos}}, \bibinfo {author} {\bibfnamefont {T.~R.}\ \bibnamefont
  {Nielsen}}, \ and\ \bibinfo {author} {\bibfnamefont {S.}~\bibnamefont
  {Bay}},\ }\bibfield  {title} {\emph {\bibinfo {title} {Fundamental quantum
  optics in structured reservoirs},\ }}\href
  {http://stacks.iop.org/0034-4885/63/i=4/a=201} {\bibfield  {journal}
  {\bibinfo  {journal} {Reports on Progress in Physics}\ }\textbf {\bibinfo
  {volume} {63}},\ \bibinfo {pages} {455} (\bibinfo {year} {2000})}\BibitemShut
  {NoStop}%
\bibitem [{\citenamefont {Shahmoon}\ \emph {et~al.}(2014)\citenamefont
  {Shahmoon}, \citenamefont {Mazets},\ and\ \citenamefont
  {Kurizki}}]{Shahmoon2014a}%
  \BibitemOpen
  \bibfield  {author} {\bibinfo {author} {\bibfnamefont {E.}~\bibnamefont
  {Shahmoon}}, \bibinfo {author} {\bibfnamefont {I.}~\bibnamefont {Mazets}}, \
  and\ \bibinfo {author} {\bibfnamefont {G.}~\bibnamefont {Kurizki}},\
  }\bibfield  {title} {\emph {\bibinfo {title} {Non-additivity in
  laser-illuminated many-atom systems},\ }}\href {\doibase
  10.1364/OL.39.003674} {\bibfield  {journal} {\bibinfo  {journal} {Opt.
  Lett.}\ }\textbf {\bibinfo {volume} {39}},\ \bibinfo {pages} {3674} (\bibinfo
  {year} {2014})}\BibitemShut {NoStop}%
\bibitem [{\citenamefont {Chang}\ \emph {et~al.}(2012)\citenamefont {Chang},
  \citenamefont {Jiang}, \citenamefont {Gorshkov},\ and\ \citenamefont
  {Kimble}}]{Chang2012a}%
  \BibitemOpen
  \bibfield  {author} {\bibinfo {author} {\bibfnamefont {D.~E.}\ \bibnamefont
  {Chang}}, \bibinfo {author} {\bibfnamefont {L.}~\bibnamefont {Jiang}},
  \bibinfo {author} {\bibfnamefont {A.~V.}\ \bibnamefont {Gorshkov}}, \ and\
  \bibinfo {author} {\bibfnamefont {H.~J.}\ \bibnamefont {Kimble}},\ }\bibfield
   {title} {\emph {\bibinfo {title} {Cavity {QED} with atomic mirrors},\
  }}\href {http://stacks.iop.org/1367-2630/14/i=6/a=063003} {\bibfield
  {journal} {\bibinfo  {journal} {New Journal of Physics}\ }\textbf {\bibinfo
  {volume} {14}},\ \bibinfo {pages} {063003} (\bibinfo {year}
  {2012})}\BibitemShut {NoStop}%
\bibitem [{\citenamefont {Fleischhauer}\ and\ \citenamefont
  {Lukin}(2000)}]{Fleischhauer2000a}%
  \BibitemOpen
  \bibfield  {author} {\bibinfo {author} {\bibfnamefont {M.}~\bibnamefont
  {Fleischhauer}}\ and\ \bibinfo {author} {\bibfnamefont {M.~D.}\ \bibnamefont
  {Lukin}},\ }\bibfield  {title} {\emph {\bibinfo {title} {Dark-state
  polaritons in electromagnetically induced transparency},\ }}\href {\doibase
  10.1103/PhysRevLett.84.5094} {\bibfield  {journal} {\bibinfo  {journal}
  {Phys. Rev. Lett.}\ }\textbf {\bibinfo {volume} {84}},\ \bibinfo {pages}
  {5094} (\bibinfo {year} {2000})}\BibitemShut {NoStop}%
\bibitem [{\citenamefont {Caneva}\ \emph {et~al.}(2015)\citenamefont {Caneva},
  \citenamefont {Manzoni}, \citenamefont {Shi}, \citenamefont {Douglas},
  \citenamefont {Cirac},\ and\ \citenamefont {Chang}}]{Caneva2015a}%
  \BibitemOpen
  \bibfield  {author} {\bibinfo {author} {\bibfnamefont {T.}~\bibnamefont
  {Caneva}}, \bibinfo {author} {\bibfnamefont {M.~T.}\ \bibnamefont {Manzoni}},
  \bibinfo {author} {\bibfnamefont {T.}~\bibnamefont {Shi}}, \bibinfo {author}
  {\bibfnamefont {J.~S.}\ \bibnamefont {Douglas}}, \bibinfo {author}
  {\bibfnamefont {J.~I.}\ \bibnamefont {Cirac}}, \ and\ \bibinfo {author}
  {\bibfnamefont {D.~E.}\ \bibnamefont {Chang}},\ }\bibfield  {title} {\emph
  {\bibinfo {title} {Quantum dynamics of propagating photons with strong
  interactions: a generalized input-output formalism},\ }}\href
  {http://stacks.iop.org/1367-2630/17/i=11/a=113001} {\bibfield  {journal}
  {\bibinfo  {journal} {New Journal of Physics}\ }\textbf {\bibinfo {volume}
  {17}},\ \bibinfo {pages} {113001} (\bibinfo {year} {2015})}\BibitemShut
  {NoStop}%
\bibitem [{\citenamefont {Mollow}(1975)}]{Mol75}%
  \BibitemOpen
  \bibfield  {author} {\bibinfo {author} {\bibfnamefont {B.~R.}\ \bibnamefont
  {Mollow}},\ }\bibfield  {title} {\emph {\bibinfo {title} {Pure-state analysis
  of resonant light scattering: {R}adiative damping, saturation, and
  multiphoton effects},\ }}\href@noop {} {\bibfield  {journal} {\bibinfo
  {journal} {Phys. Rev. A}\ }\textbf {\bibinfo {volume} {12}},\ \bibinfo
  {pages} {1919} (\bibinfo {year} {1975})}\BibitemShut {NoStop}%
\bibitem [{\citenamefont {Chang}\ \emph {et~al.}(2011)\citenamefont {Chang},
  \citenamefont {Safavi-Naeini}, \citenamefont {Hafezi},\ and\ \citenamefont
  {Painter}}]{Chang2011a}%
  \BibitemOpen
  \bibfield  {author} {\bibinfo {author} {\bibfnamefont {D.~E.}\ \bibnamefont
  {Chang}}, \bibinfo {author} {\bibfnamefont {A.~H.}\ \bibnamefont
  {Safavi-Naeini}}, \bibinfo {author} {\bibfnamefont {M.}~\bibnamefont
  {Hafezi}}, \ and\ \bibinfo {author} {\bibfnamefont {O.}~\bibnamefont
  {Painter}},\ }\bibfield  {title} {\emph {\bibinfo {title} {Slowing and
  stopping light using an optomechanical crystal array},\ }}\href
  {http://stacks.iop.org/1367-2630/13/i=2/a=023003} {\bibfield  {journal}
  {\bibinfo  {journal} {New Journal of Physics}\ }\textbf {\bibinfo {volume}
  {13}},\ \bibinfo {pages} {023003} (\bibinfo {year} {2011})}\BibitemShut
  {NoStop}%
\bibitem [{\citenamefont {Chang}\ \emph {et~al.}(2007)\citenamefont {Chang},
  \citenamefont {Sorensen}, \citenamefont {Demler},\ and\ \citenamefont
  {Lukin}}]{Chang2007a}%
  \BibitemOpen
  \bibfield  {author} {\bibinfo {author} {\bibfnamefont {D.~E.}\ \bibnamefont
  {Chang}}, \bibinfo {author} {\bibfnamefont {A.~S.}\ \bibnamefont {Sorensen}},
  \bibinfo {author} {\bibfnamefont {E.~A.}\ \bibnamefont {Demler}}, \ and\
  \bibinfo {author} {\bibfnamefont {M.~D.}\ \bibnamefont {Lukin}},\ }\bibfield
  {title} {\emph {\bibinfo {title} {A single-photon transistor using nanoscale
  surface plasmons},\ }}\href {http://dx.doi.org/10.1038/nphys708} {\bibfield
  {journal} {\bibinfo  {journal} {Nat Phys}\ }\textbf {\bibinfo {volume} {3}},\
  \bibinfo {pages} {807} (\bibinfo {year} {2007})}\BibitemShut {NoStop}%
\bibitem [{\citenamefont {Hafezi}\ \emph {et~al.}(2012)\citenamefont {Hafezi},
  \citenamefont {Chang}, \citenamefont {Gritsev}, \citenamefont {Demler},\ and\
  \citenamefont {Lukin}}]{Hafezi2012a}%
  \BibitemOpen
  \bibfield  {author} {\bibinfo {author} {\bibfnamefont {M.}~\bibnamefont
  {Hafezi}}, \bibinfo {author} {\bibfnamefont {D.~E.}\ \bibnamefont {Chang}},
  \bibinfo {author} {\bibfnamefont {V.}~\bibnamefont {Gritsev}}, \bibinfo
  {author} {\bibfnamefont {E.}~\bibnamefont {Demler}}, \ and\ \bibinfo {author}
  {\bibfnamefont {M.~D.}\ \bibnamefont {Lukin}},\ }\bibfield  {title} {\emph
  {\bibinfo {title} {Quantum transport of strongly interacting photons in a
  one-dimensional nonlinear waveguide},\ }}\href {\doibase
  10.1103/PhysRevA.85.013822} {\bibfield  {journal} {\bibinfo  {journal} {Phys.
  Rev. A}\ }\textbf {\bibinfo {volume} {85}},\ \bibinfo {pages} {013822}
  (\bibinfo {year} {2012})}\BibitemShut {NoStop}%
\bibitem [{\citenamefont {Fleischhauer}\ and\ \citenamefont
  {Lukin}(2002)}]{Fleischhauer2002}%
  \BibitemOpen
  \bibfield  {author} {\bibinfo {author} {\bibfnamefont {M.}~\bibnamefont
  {Fleischhauer}}\ and\ \bibinfo {author} {\bibfnamefont {M.~D.}\ \bibnamefont
  {Lukin}},\ }\bibfield  {title} {\emph {\bibinfo {title} {Quantum memory for
  photons: Dark-state polaritons},\ }}\href {\doibase
  10.1103/PhysRevA.65.022314} {\bibfield  {journal} {\bibinfo  {journal} {Phys.
  Rev. A}\ }\textbf {\bibinfo {volume} {65}},\ \bibinfo {pages} {022314}
  (\bibinfo {year} {2002})}\BibitemShut {NoStop}%
\bibitem [{\citenamefont {Hung}\ \emph {et~al.}(2013)\citenamefont {Hung},
  \citenamefont {Meenehan}, \citenamefont {Chang}, \citenamefont {Painter},\
  and\ \citenamefont {Kimble}}]{Hung2013a}%
  \BibitemOpen
  \bibfield  {author} {\bibinfo {author} {\bibfnamefont {C.-L.}\ \bibnamefont
  {Hung}}, \bibinfo {author} {\bibfnamefont {S.~M.}\ \bibnamefont {Meenehan}},
  \bibinfo {author} {\bibfnamefont {D.~E.}\ \bibnamefont {Chang}}, \bibinfo
  {author} {\bibfnamefont {O.}~\bibnamefont {Painter}}, \ and\ \bibinfo
  {author} {\bibfnamefont {H.~J.}\ \bibnamefont {Kimble}},\ }\bibfield  {title}
  {\emph {\bibinfo {title} {Trapped atoms in one-dimensional photonic
  crystals},\ }}\href {http://stacks.iop.org/1367-2630/15/i=8/a=083026}
  {\bibfield  {journal} {\bibinfo  {journal} {New Journal of Physics}\ }\textbf
  {\bibinfo {volume} {15}},\ \bibinfo {pages} {083026} (\bibinfo {year}
  {2013})}\BibitemShut {NoStop}%
\end{thebibliography}

%

\end{document}